\documentclass[11pt,a4paper]{article}
\usepackage[utf8]{inputenc}
\usepackage[T1]{fontenc}
\usepackage{mathptmx} 
\usepackage[pdftex]{graphicx} 
\usepackage{geometry}
\usepackage[english]{babel} 
\usepackage{calc} 
\usepackage{enumitem} 
\usepackage[]{biblatex}

\DeclareNameAlias{author}{last-first}
\DeclareDelimFormat[bib]{multinamedelim}{\addcomma\space}
\DeclareDelimFormat[bib]{finalnamedelim}{\addcomma\space}
\DeclareDelimFormat[parencite]{nameyeardelim}{\addcomma\space}
\usepackage{amsmath}
\usepackage{amsfonts}
\usepackage{amssymb}
\usepackage{xspace}
\usepackage{lmodern}
\usepackage{algorithm}
\usepackage{algpseudocode}
\usepackage{rotating}
\usepackage{tikz}
\usepackage{placeins}
\usetikzlibrary{decorations}
\usepackage{csquotes}
\usepackage{caption}
\usepackage{subcaption}
\usepackage{amsthm}
\usepackage{appendix}
\usepackage[pdftex,linkcolor=black,pdfborder={0 0 0}]{hyperref}
\usepackage{hyperref}
\usepackage{caption}
\usepackage{tabularx}
\setlist[enumerate]{itemsep=0mm}

\newcommand{\Ex}{\mathbb{E}}
\newcommand{\V}{\mathbb{V}}

\newcommand{\R}{\mathbb{R}}

\newcommand{\F}{{\mathcal F}}

\newcommand{\Qm}{Q_{\text{main}}}
\def\orcid#1{\kern .08em\href{https://orcid.org/#1}{\includegraphics[keepaspectratio,width=0.7em]{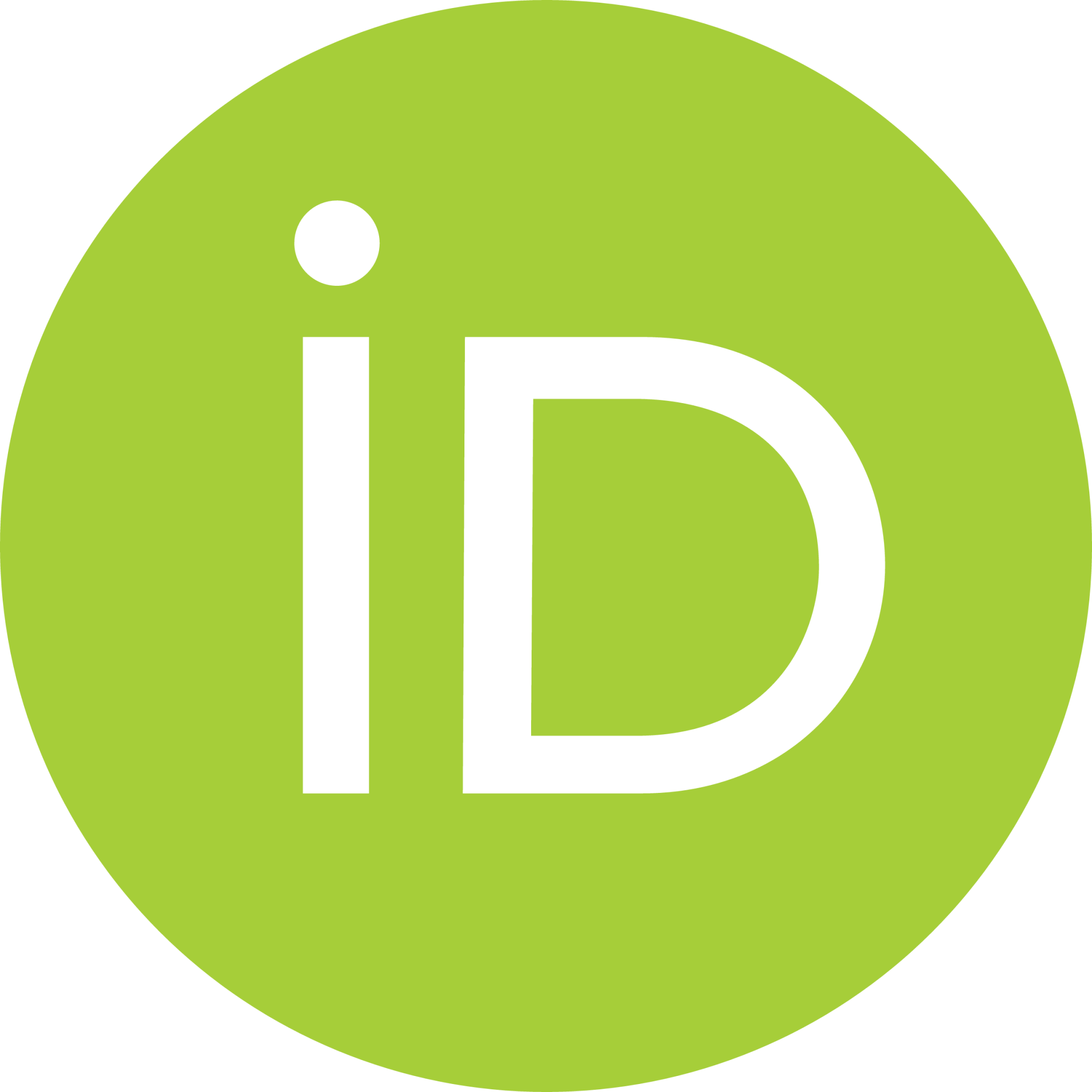}}}

\DeclareMathOperator*{\argmax}{arg\,max}
\DeclareMathOperator*{\argmin}{arg\,min}

\newtheorem{thm}{Theorem}
\newtheorem{prop}{Proposition}
\newtheorem{defi}{Definition}
\newtheorem{remark}{Remark}

\hypersetup{
  pdfsubject = {MultiAgent Optimal execution with RL},
  pdftitle = {Tacit Collusion and deviations from Nash equilibrium},
  pdfauthor = {Andrea Macrì and Fabrizio Lillo}
}

\begin{document}

\title{{Deviations from the Nash equilibrium in a two-player optimal execution game with reinforcement learning}}

\author{Fabrizio Lillo$^{1,2}$\footnote{fabrizio.lillo@sns.it} \orcid{0000-0002-4931-4057} and Andrea Macrì$^1$\footnote{andrea.macri@sns.it} \orcid{0000-0001-6526-5884} \\
\small{$^1$ Scuola Normale Superiore, Pisa, Italy}\\
\small{$^2$ Dipartimento di Matematica University of Bologna, Bologna, Italy}}

\date{\today}
\maketitle
\begin{abstract}
The use of reinforcement learning algorithms in financial trading is becoming increasingly prevalent. However, the autonomous nature of these algorithms can lead to unexpected outcomes that deviate from traditional game-theoretical predictions and may even destabilize markets. In this study, we examine a scenario in which two autonomous agents, modelled with Double Deep Q-Learning, learn to liquidate the same asset optimally in the presence of market impact, under the Almgren-Chriss (2000) framework. We show that the strategies learned by the agents deviate significantly from the Nash equilibrium of the corresponding market impact game. Notably, the learned strategies exhibit supra-competitive solution, {which might be compatible with a tacit collusive behaviour}, closely aligning with the Pareto-optimal solution. We further explore how different levels of market volatility influence the agents' performance and the equilibria they discover, including scenarios where volatility differs between the training and testing phases.
\end{abstract}

\section{Introduction}

The increasing automation of trading over the past three decades has profoundly transformed financial markets. The availability of large, detailed datasets has facilitated the rise of algorithmic trading—a sophisticated approach to executing orders that leverages the speed and precision of computers over human traders. Using diverse data sources, these automated systems have revolutionized the way trades are conducted. In 2019, it was estimated that approximately 92\% of the trading in the foreign exchange (FX) market was driven by algorithms \cite{kissell2020algorithmic}. The rapid advancements in Machine Learning (ML) and Artificial Intelligence (AI) have significantly accelerated this trend, leading to an increasing adoption of autonomous trading algorithms {and, as argued in \cite{azzutti2021machine}, this poses problems from a legislative point of view when it comes to deciding whether and how regulators can hold accountable firms for malpractices due to algorithms' misuse. 
In the landscape of autonomous trading systems, increasing interest, by both practitioners and academics alike, is being shown on those based on Reinforcement Learning (RL), especially deep RL. Deep RL algorithms differ from traditional supervised learning models. }Instead of being trained on labelled input/output pairs, RL algorithms explore a vast space of potential strategies, learning from past experiences to identify those that maximise long-term rewards. This approach allows for the continuous refinement of trading strategies, further enhancing the efficiency and effectiveness of automated trading.

The flexibility of {deep RL} comes with significant potential costs, particularly due to the opaque, black-box nature of these algorithms. This opacity can lead to unexpected outcomes that may destabilize the system they control—in this case, financial markets—through the actions they take, such as executing trades. The complexity and risk increase further when multiple autonomous agents operate simultaneously within the same market. The lack of transparency in RL algorithms may result in these agents inadvertently learning joint strategies that deviate from the theoretical Nash equilibrium, potentially leading to unintended market manipulation. Among the various risks, the emergence of collusion is particularly notable. Even without explicit instructions to do so, RL agents may develop cooperative strategies that manipulate the market, a phenomenon known as {\it tacit collusion}. This type of emergent behaviour is especially concerning because it can arise naturally from the interaction of multiple agents, posing a significant challenge to market stability and fairness.

In this paper, we investigate the equilibria and the emergence of
{supra-competitive equilibria and their implications in terms of tacit collusive behaviour} in a market where two autonomous agents, {modelled} by deep RL, engage in an optimal execution task. Although tacit collusion in RL settings has been explored in areas such as market making (as reviewed below), much less is known about how autonomous agents behave when trained to optimally trade a large position. We adopt the well-established Almgren-Chriss framework first introduced in \cite{almgren}, for modelling market impact, focusing on two agents tasked with liquidating large positions of the same asset. The agents are modelled using Double Deep Q-Learning (DDQL) and engage in repeated trading to learn the optimal equilibrium of the associated open-loop game.

For this specific setup, the Nash equilibrium of the game has been explicitly derived in \cite{schied2017state}, providing a natural benchmark against which we compare our numerically derived equilibria. In addition, we explicitly derive the Pareto-optimal strategy and numerically characterize the Pareto-efficient set of solutions. Our primary goal is to determine whether the two agents, without being explicitly trained to cooperate or compete, can naturally converge to a non Nash equilibrium, when it  existence and uniqueness is known.

Our findings reveal that the strategies learned by the RL agents deviate significantly from the Nash equilibrium of the market impact game. Across various levels of market volatility, we observe that the learned strategies exhibit {seemingly tacit collusive behaviour}, closely aligning with the Pareto-optimal strategy. Given that financial market volatility is time-varying, we also examine the robustness of these strategies when trained and tested under different volatility regimes. Remarkably, we find that the strategies learned in one volatility regime remain supra-competitive, even when adopted in various volatility conditions, underscoring the robustness of the training process.

\paragraph{Literature review.} Optimal execution has been extensively studied in the financial literature. Starting from the seminal contributions in \cite{bertsimas} and in \cite{almgren}, many authors have contributed to extend the model's consistency with reality. {Notable examples in this area are \cite{cartea2015algorithmic,cartea2016incorporating,casgrain2019trading} for optimal control perspective, \cite{bouchaud2003fluctuations,bouchaud2009markets,gueant2012optimal,gatheral2012transient,obizhaeva2013optimal} for the transient impact extensions and \cite{gueant2015general} for a limit order generalisation of the Almgren and Chriss framework.} In its basic setting, the optimal execution problem considers just one agent unwinding or acquiring a quantity $q_0$ of assets within a certain time window. However, if many other agents are considered to be either selling or buying, thus pursuing their own optimal execution schedules in the same market, the model becomes more complicated since it now requires modelling other agents' behaviour the same market. 
Using a large dataset of real optimal executions, \cite{bucci2020co} shows that the cost of an execution strongly depends on the presence of other agents liquidating the same asset.
The increased complexity when treating the problem in this way allows for more consistency with reality and, at the same time, opens the path for further studies on how the agents interact with each other in such a context.
\par
The optimal execution problem with $n$-agents operating in the same market has thus been studied under a game theoretic lens in both its closed-loop (in \cite{carmona2011predatory, micheli2023closed}) and open-loop (in \cite{brunnermeier2005predatory, carlin2007episodic, schoneborn2009liquidation, schied2017state, schied2019market}) formulations. In more recent works, rather than just optimal execution, liquidity competition is also analysed (as in \cite{drapeau2019fbsde,neuman2023trading}) along with market making problems (as in \cite{cont2021interbank, cont2022dynamics}).
\par
In recent times, with the advancements of machine learning techniques, the original optimal execution problem has been extensively studied using RL techniques. Among the many\footnote{For a more comprehensive overview of the state of the art on RL methods in finance, please refer to \cite{sun2023reinforcement} and \cite{hambly2023recent}}, some examples in the literature on RL techniques applied to the optimal execution problem are found in \cite{ning2021double, schnaubelt2022deep, macrì2024reinforcement}.
\par
Applications of multi-agent RL to financial problems are {scarce} if compared to those that study the problem in a single agent scenario. Optimal execution in a multi-agent environment is tackled in \cite{karpe2020multi}, where the authors analyse the optimal execution problem form the standpoint of a many agents environment and an RL agent trading on Limit Order Book. The authors test and develop a multi-agent Deep Q-learning algorithm able to trade on both simulated and real order books data. Their experiments converge to agents who adopt only the so-called Time Weighted Average Price (TWAP) strategies in both cases. On the other hand, still on optimal execution problem's applications of multi-agent RL, the authors in \cite{bao2019multi} analyse the interactions of many agents on their respective optimal strategies under the standpoint of cooperative competitive behaviours, adjusting the reward functions in the Deep Deterministic Policy Gradient algorithm used, in order to allow for either of the two phenomena, but still using the basic model introduced in \cite{almgren}, and modelling the interactions of the agents via full reciprocal disclosure of their rewards and not considering the sum of the strategies to be a relevant feature for the permanent impact in the stock price dynamics. \par

The existence of collusion and the emergence of collusive behaviours are probably the most interesting phenomena based on agents interaction in a market, as these are circumstances that might naturally arise in markets even if no instruction on how to and whether to collude or not, are given to the modelled agents. 

{One of the first contributions to collusion theory of autonomous algorithms has been put forth in~\cite{waltman2008q}. In this contribution the authors show empirically by means of Q-learning, how two firms in a production game with linear demand function, can learn collusive behaviour even if no direct communication and explicit punishment strategies are present. Hence, in this contribution the authors show how competing producers in a Cournot oligopoly learn to increase prices above Nash equilibrium by reducing production, so that firms learn to collude even if they do not learn to obtain the highest possible joint profit. Thus, the firms reach a supra-competitive equilibrium, comparable to the collusive one even if no punishment mechanism is explicitly present or direct communication between the players is present. This remarkable result has been further analysed and extended in~\cite{calvano2020artificial}. Here, the authors show how using Q-learning, two agents in a duopoly are able to retrieve supra-competitive equilibria without information exchange. The equilibrium is sustained by collusive strategies enforced by eventual punishment phases in the case where an agent deviates from said collusive equilibrium. 
\\
\\
Notwithstanding the empirical emergence of collusive behaviour, some concerns have emerged on whether or not the supra-competitive outcomes of such games are indeed collusions or are due to some algorithm-specific features and thus, if these possibly tacit collusive behaviours have to be a major concern for the regulators. 
This topic has been investigated in \cite{abada2023artificial}, where the authors consider battery and electricity markets to empirically prove how such agents would indeed punish \emph{any} deviation from a collusive strategy by inducing shocks to the environment where the agents are playing after they have converged in the training phase. They conjecture that this seeming collusion is due to imperfect exploration of states and parameter initialisation. 
More recently,  \cite{abada2024collusion} expands this point of view by showing how in a Q-learning environment seemingly supra-competitive equilibria are due to the poor exploration policy of unsophisticated algorithms. They show how for simple Q-learning algorithms there is an auto-induced cooperative trap if states are not thoroughly explored and learning rate decreases `too fast', consequently implying that cooperative strategies have not been explored enough during training. This is opposite to what the authors find with more sophisticated algorithms where collusive behaviours do not emerge. The conclusion is that the supra-competitive outcomes should be of concern for market authorities if and only if the algorithms employed are not sophisticated.
Also Ref. \cite{lambin2024less} investigates when supra-competitive equilibria result necessarily from collusive strategies based on reward and punishment mechanisms to sustain cooperation. Unlike previous studies, their analysis critically suggests that both high prices and these strategic patterns actually emerge from a combination of simultaneous exploration and learning inertia intrinsic to reinforcement learning techniques, without any direct causal connection. They show that such forms of apparent collusion can arise even in memoryless settings and with myopic\footnote{Agents with a low discount factor on future rewards, i.e. that value more immediate rewards over future possible ones.} agents. Based on these findings, they caution against making straightforward analogies with human collusion and suggest addressing elevated algorithmic prices through tailored regulatory measures.}

{There is no unanimous consensus on whether algorithmic collusion does --in fact-- naturally emerge from the interaction of autonomous agents, and on the drivers that lead to supra-competitive equilibria. Especially, because of the definition of collusion\footnote{“Collusion is when firms use strategies that embody a reward-punishment scheme
which rewards a firm for abiding by the supra-competitive outcome and punishes it for
departing from it” as defined in \cite{harrington2018developing}.} that, according to \cite{abada2023artificial} is more suitable to a human to human cooperative behaviour rather than to an algorithmic driven phenomenon. 
Thus, a part of the literature points out at these phenomena as tacitly collusive equilibria while another stream interprets these results as not necessarily collusive even if the 
strategies employed by the agents might suggest so. This implies that applying existing regulatory 
solutions to markets where prices are set through reinforcement learning algorithms may miss the point, and thus interventions by the regulator on such markets might eventually be inadequate. In this fragmented landscape, literature on collusion in financial markets, to the best of our knowledge, has just started to develop.
}
In fact, for what concerns financial markets, the problem becomes more involved since the actors in the market do not directly set the price in a `one sided' way as is the case for production economies that are studied in economic literature. Among the many contributions that study the emergence of collusive behaviours in financial markets, in \cite{xiong2022interactions} it is shown how tacit collusion arises between market makers, modelled using deep RL, in a competitive market. In \cite{cartea2022algorithmic} the authors show how market making algorithms tacitly collude to extract rents, and this behaviour strictly depends on tick size and coarseness of price grid. Finally, in \cite{cont2022dynamics} the authors use a multi-agent deep RL algorithm to model market makers competing in the same market, the authors show how competing market makers learn how to adjust their quotes, giving rise to tacit collusion by setting spread levels strictly above the competitive equilibrium level. 

{To the best of our knowledge no prior work did consider supra-competitive, possibly tacit collusive, equilibria in optimal execution problems. As discussed in the Conclusions section, the existence of such equilibria might indeed imply a loss of welfare for other market participants. Therefore, studying the behaviour of learning agents opens for rather new perspectives on how collusion eventually arises through learning. Finally, studying supra-competitive equilibria and potential collusion for this class of problems is thus instrumental to understanding, and eventually, easing the regulatory tasks of market authorities when it comes to assess possible informational distortions of assets prices which, in turn, might affect the correct functioning of financial markets, resulting in welfare loss for market participants. 
This paper aims at tackling these points, while enlarging the analysis on these equilibria to the optimal execution problem.
}

The paper is organised as follows: Section 2 introduces the theoretical setting of the market impact game, Section 3 introduces the DDQL algorithm for the multi-agent optimal execution problem, Section 4 discusses the results for different parameter settings, {in Section 5 we critically discuss whether the observed results are tacit collusion or supra-competitive outcomes and finally Section 6 provides conclusions and outlines further research directions}.
\par

\section{Market impact game setting}
\label{sec:mkt_set}

\paragraph{The Almgren-Chriss model.} The setup of our framework is based on the seminal Almgren-Chriss model first introduced in \cite{almgren} for optimal execution. In this setting, a single agent wants to liquidate an initial inventory of $q_0$ shares within a time window $[0, T]$, which, in the discrete-time setting, is divided into $N$ equal time increments of length $\tau = T/N$. The main assumption of the model is that the mid-price (or the efficient price) evolves according to a random walk with a drift depending on the traded quantity. Moreover, the price obtained by the agent differs from the mid-price by a quantity that depends on the volume traded in the interval. More formally, let $S_t$ and $\tilde S_t$ be the mid-price and the price received by the agent at time $t$, and let $v_t$ be the number of shares traded by the agent in the same interval; then the dynamics is given by

\begin{align}
\begin{split}
    \label{eq:ABM_AeC}
    {S}_{t} &= S_{t-1} - {p}\left( {v_t}/{\tau}\right)\tau+\sigma\tau^{\frac{1}{2}}\xi_t\\
    \Tilde{S_t} &=  {S}_{t-1} - h\left( {v_t}/{\tau}\right) {\quad .}
\end{split}
\end{align}
$S_t$ evolves because of a diffusion part {$ \xi_t\sim\mathcal{N}(0, 1) $} multiplied by the price volatility $\sigma$ and a drift part, termed \emph{permanent impact}, ${p}\left( {v_t}/{\tau}\right)$, assumed to be linear and constant: ${p}\left( {v_t}/{\tau}\right) = \kappa {v_t}/{\tau}$. The price  \(\Tilde{S}_t\) received by the agent is equal to the mid-price $S_t$ but impacted by a \emph{temporary impact} term also assumed to be linear and constant in time: \(h(v_t/{\tau}) = \alpha{v_t}/{\tau}\).

{These two impacts are the classic components of market impact. In fact, they represent the adverse movement of the asset price generated by trades. Permanent impact reflects permanent changes in the asset's fundamental price process, while the temporary impact should reflect some temporary market imbalances. The former is cumulatively assumed to impact the asset price at least until the end of the execution window, while the latter vanishes as soon as the trade has been performed.}

The aim of the agent is to unwind their initial portfolio maximising the cash generated by their trading over the $N$ time steps. This objective can be rewritten in terms of Implementation Shortfall (IS)\footnote{{Notice that the IS could, in principle, take negative values as well. But in the absence of exogenous price drift and linear impact functions, as in our case, its expected value is positive.}} that is defined in the single agent case as:
\begin{equation}
\label{eq:is}
    IS(\Vec{v}) = S_0q_0 - \sum^N_{t=1} \Tilde{S}_tv_t {\quad ,}
\end{equation}
where $\Vec{v}=(v_1,...,v_N)$ 
is the vector that contains the positive quantities sold at each time step, i.e. $v_t \geq 0,~ \forall~ t \in [1,N]$ and \(S_0\) is the mid-price at the beginning of the trading window. The optimisation problem faced by the agent can be written as 
\begin{equation}
\label{eq:mean_var}
    \min_{\Vec{v}} \,\,\Ex[IS(\Vec{v})] {+} {\lambda}\V[IS(\Vec{v})]~~~~~s.t.~~\sum_{t=1}^Nv_t=q_0 {\quad ,}
\end{equation}
where $\lambda$ is the risk aversion parameter of the agent, {\(\Ex[\cdot]\) and \(\V[\cdot]\) refer to the expected value and the variance of the IS functional respectively.} Under the linearity assumption of the two impacts, the problem can be easily solved analytically. In the following, we are going to consider risk neutral agents ($\lambda=0$) and in this case the optimal trading schedule is the TWAP, i.e. $v_t=q_0/N$, where the trading velocity is constant.  

\paragraph{Almgren-Chriss market impact game.} Now we consider two agents selling an initial inventory $q_0$ shares within the same time window time window $[0, T]$. The traded quantities by the two agents in the time step $t$ are indicated with $v^{(1)}_t$ and $v^{(2)}_t$  and $V_t = v^{(1)}_t + v^{(2)}_t$ is the total quantity traded in the same interval. The equations for the dynamics in discrete time are 
\begin{align}
\begin{split}
    \label{eq:perm}
    {S}_{t} &= S_{t-1} - {p}\left( {V_t}/{\tau}\right)\tau+\sigma\tau^{\frac{1}{2}}\xi_t\\
    ~~\Tilde{S_t}^{(k)} &=  {S}_{t-1} - h\left( {v^{(k)}_t}/{\tau}\right)~~~~~k=1,2 {\quad ;}
\end{split}
\end{align}
i.e. the mid-price is affected by the total traded volume $V_t$, while the price received by each agents depends on the quantity they trade.
{Notice that, according to \cite{schied2017state} the temporary impact depends only on the volume traded by the agent singularly and not on the volume traded by the other agent. This corresponds to the case where the asset is relatively liquid. Although it would be interesting to extend the model to the case of a temporary impact that depends on both agents' traded volumes, in the present work we will consider the case in Eq.~\eqref{eq:perm}.}
Since more than one agent are optimising their trading volumes and the cash received depends on the trading activity of the other agent, the natural setting to solve this problem is the one of game theory. Since each agent is not directly aware of the selling activity of the other agent and  the two agents interact through their impact on the price process, the resulting problem is an open-loop game. The existence and uniqueness of a Nash equilibrium in such a symmetric open-loop game has been studied in \cite{schied2017state} in the case of \emph{n} different players and linear impact functions. First of all, they defined the Nash equilibrium, {in continuous time and considering the vector of levels of inventory held at each time-step $t$, i.e. $\vec{q}^{~(k)}$}, as: 
\begin{defi}[Nash equilibrium \cite{schied2017state}]
    In an $n$ players game, where $n \in \mathbb{N}$, with $q^{(1)}_0, \dots, q^{(n)}_0 \in \R $ initial inventory holdings and $\lambda_1, \dots, \lambda_n$ non-negative coefficients of risk aversion.
    A Nash equilibrium for mean-variance optimisation as in Eq.~\eqref{eq:mean_var} is a collection of $\mathbf{q^*} = \{ \Vec{q}^{\,(1)^*}, \dots, \Vec{q}^{\,(n)^*}\}$ inventory holdings such that for each $k \in n, ~ {\Vec{q}^{~(k)}} \in \mathcal{A}_{\text{det}}$ the mean variance functional is minimised:
    \begin{equation*}
         \Ex[IS(\Vec{q}^{\,(k)}| \mathbf{q^{\,(-k)^*}})] + {\lambda_k}\V[IS(\Vec{q}^{\,(k)}| \mathbf{{q}^{\,(-k)^*}})]
    \end{equation*}
    for each agent $k$,\footnote{Where $\mathcal{A}_{\text{det}}$ is the set of all admissible deterministic trading strategies, for a more formal definition see \cite{schied2017state}.} where $ \mathbf{{q}^{(-k)^*}}$ are the strategies of the other players minus the strategy of the $k$-agent that is being considered.
\end{defi}

{The set of deterministic strategies, $\mathcal{A}_{\text{det}}$, in the definition is the set of the strategies that is not adapted to the mid-price level. Thus, it does not depend on the realisation of the noise \(\xi_t\) in the dynamics of the price process.
}

Ref. \cite{schied2017state} proves that there exists a unique Nash equilibrium for the mean-variance optimisation problem. The unique Nash equilibrium strategy $q^{(k)^*}_t$, the Nash remaining inventory of agent $k$ at time $t$, for $n$ players is given as a solution to the second-order system of differential equations:
\begin{equation}
    \lambda^{(k)}\sigma^{2}q^{(k)}_t - 2{\alpha}\tau\Ddot{\,q}^{(k)}_t = \kappa\sum_{j\neq k}\Dot{q}^{\,(k)}_t + {\alpha}\tau\sum_{j\neq k}\Ddot{\,q}^{(k)}_t
\end{equation}
with two-point boundary conditions
\begin{equation}
    q^{(k)}_0 = q_0 \hspace{0.5cm}\text{and}\hspace{0.5cm}q^{(k)}_T=0 \,\, , \hspace{0.5cm}\forall\,\, k=1,\dots,n
\end{equation}
Focusing on the special case of two players, in \cite{schied2017state} it is proved that the selling schedule at the unique Nash equilibrium is:
\begin{equation}
    \label{eq:opt_inv_two_players}
    q^{(1)^*}_t = \frac12 (\Sigma(t) + \Delta(t))\hspace{2cm}\text{and}\hspace{2cm}q^{(2)^*}_t = \frac12 (\Sigma(t) - \Delta(t))\,,
\end{equation}
with:
\begin{align}
\begin{split}
    \label{eq:meaning}
    \Sigma(t) = Q e^{-\frac{\kappa t }{6\alpha}} \frac{\sinh{\left( \frac{(N - t) \sqrt{\kappa^2 + 12\alpha\lambda\sigma^2}}{6\alpha} \right)}}{\sinh{\left( \frac{N\sqrt{\kappa^2+12\alpha\lambda\sigma^2}}{6\alpha}\right)}}
\hspace{1cm}\text{and}\hspace{1cm}
    \Delta(t) = \tilde{Q} e^{\frac{\kappa t }{2\alpha}} \frac{\sinh{\left( \frac{(N - t) \sqrt{\kappa^2 + 4\alpha\lambda\sigma^2}}{2\alpha} \right)}}{\sinh{\left( \frac{N\sqrt{\kappa^2+4\alpha\lambda\sigma^2}}{2\alpha}\right)}}
\end{split}
\end{align}
where $Q = q^{(1)}_0 + q^{(2)}_0 $ and $\Tilde{Q} = q^{(1)}_0 - q^{(2)}_0 $.
It can be noticed that the Nash inventory holdings, and thus the trading rates, now depend also on the permanent impact $\kappa$, contrary to what happens in the single agent case studied in \cite{almgren}. 
The price level does not enter directly into the optimal inventory formula, but the permanent impact and the volatility of the asset on the other hand do, proportionally to agents' risk aversion.
\begin{remark}
{
In \cite{schied2017state} the authors derive the unique Nash equilibrium considering the continuous time version of Eq.\eqref{eq:perm}. 
We will use the Nash inventory holdings and the selling strategy associated to Eq.\eqref{eq:opt_inv_two_players}-\eqref{eq:meaning} discretised over time. We choose to consider the discretised version of the Nash equilibrium, and to derive results in discrete time as they are functional to the numerical and deep RL experiments, as well as to highlight the differences that might arise when autonomous agents are trained together in a similar but subtly different environment. The Nash equilibrium in the discrete time setting was derived in Sec. 2 of \cite{cordoni2024transient}, considering a constant decay kernel in a transient impact market.\footnote{In \cite{schied2017high}the authors show that the high-frequency limits of the discrete-time equilibrium costs converge to the expected
costs in the continuous-time Nash equilibrium.}
}
\end{remark}
\paragraph{Beyond the Nash equilibrium.} 

{In this setting, as shown a \textit{unique} Nash equilibrium exists but, leveraging on reviewed literature, other non-Nash equilibria might as well. Especially, supra-competitive and collusive equilibria have been shown to exists in oligopolistic settings such as the one considered in a market impact game. To this end, we aim at studying the behaviour of two agents who have to solve an optimal execution problem in a multi-agent setting. We study the problem from a \textit{learning} perspective, meaning that we will use deep RL to train the agents to solve the liquidation problem in such setup; we then study the strategies found by the agents, comparing the cost of the learnt strategies with the ones prescribed by both the Nash equilibrium strategy and the Pareto optimal strategy, that will be shown to be the collusive strategy. In order to do so, we focus on the case where two {\it risk neutral} agents want to liquidate the same initial portfolio made by an amount $q_0$ of the same asset. In the numerical setup we let the agents play multiple episode instances of the optimal execution problem. This means that, defining a trading episode to be the complete unwinding of the initial inventory $q_0$ over time $[0,T]$ in $N$ time steps by both agents, the overall game is made by repeated trades at each $t=1,\dots,N$ time-steps and by $B$ iterations of trading, each containing a full inventory liquidation episode. Therefore, for each iteration $i$ we will consider two vectors $\vec{v}^{\,(1)}$ and $\vec{v}^{\,(2)}$ containing the trading schedule for the two agents, .}

{The numerical setup is justified by the episodical nature of the game under consideration, as by definition -and differently from other setups- the optimal execution problem happens over a finite time-window. Each vector is associated with its IS cost, according to Eq. \ref{eq:is}. This cost, along with the strategies' vectors are going to be the main objective of our study. We now define the set of admissible strategies and the average IS cost.}

\begin{defi}[]
The set $\mathcal{A}$ of admissible selling strategies 
is composed by the pair of vectors $\Vec{v}^{\,(1,2)} = (\vec{v}^{\,(1)}, \vec{v}^{\,(2)})$, such that for $k = 1,2$:
        \begin{itemize}
            \item $\sum^N_{t=1}v^{(k)}_{{t}}=q_0$
            \item $\Vec{v}_{{t}}^{(k)}$ is $(\F_t)_{t\geq0}$ - adapted and progressively measurable
            \item $\sum^N_{t=1}(v^{(k)}_{{t}})^2<\infty$
        \end{itemize}
For any $\Vec{v}^{\,(1,2)} \in \mathcal{A}$ the average Implementation Shortfall is defined as:
\begin{equation}
   \Bar{IS}\left(\Vec{v}^{\,(1,2)}\right) = \frac{1}{N}\sum^N_{t=1} IS\left(\Vec{v}^{\,(1,2)}_{{t}}\right)\,.
\end{equation}
where $IS\left(\Vec{v}^{\,(1,2)}_{{t}}\right) = IS\left(\Vec{v}^{\,(1)}_{{t}}\right) + IS\left(\Vec{v}^{\,(2)}_{{t}}\right)$.
\end{defi}

Leveraging on \cite{cont2022dynamics} and \cite{cont2021interbank}, we define \textit{collusive} strategies and we show that they are necessarily Pareto-optimal.
{Notice that, given the iterated nature of the game due to the time and inventory constraints that -by design- constrain the agents, we will consider the following equilibrium properties to hold over several iterations of the game. Although the same properties would hold in a one-shot game, we aim at studying what equilibrium would naturally arise when the agents are simultaneously \textit{learning}, without being told to either cooperate or compete. The theoretical results stated below are functional to the analysis of the learned equilibria, therefore in the numerical experiments part we depart from a one-shot game setup in favour of an iterated game setup.}

\begin{defi}[Collusion]
\label{defi:collusion} 
   A pair of vectors of selling schedules $\Vec{v}^{\,(1,2)}_c = (\vec{v}^{\,(1)}_c, \vec{v}^{\,(2)}_c)\,\, \in \mathcal{A}$ is defined to be a collusion if  
   $\forall\,\, \Vec{v}^{\,(1,2)} \in \mathcal{A}$:
    \begin{equation*}
        \Bar{IS}(\Vec{v}^{\,(1,2)}_c) \leq \Bar{IS}(\Vec{v}^{\,(1,2)})
    \end{equation*}
\end{defi}

\begin{defi}[Pareto Optimum]
    \label{defi:pareto}
    A pair of vectors of selling schedules $\Vec{v}^{\,(1,2)}_p = (\vec{v}^{\,(1)}_p, \vec{v}^{\,(2)}_p)\,\, \in \mathcal{A}$ is a Pareto optimal strategy if and only if there does not exist $\Vec{v}^{\,(1,2)}\in\mathcal{A}$ such that:
    \begin{align}
        \begin{split}
            \forall\,\,i=1,\dots,B\,, \,~~~& \Bar{IS}(\Vec{v}^{\,(1,2)}_p)\geq\Bar{IS}(\Vec{v}^{\,(1,2)})\\
            \exists\,\,j=1,\dots,B\,, ~~~& \Bar{IS}(\Vec{v}^{\,(1,2)}_p)>\Bar{IS}(\Vec{v}^{\,(1,2)})\\
        \end{split}
    \end{align}
\end{defi}
\begin{prop}[Collusive Pareto optima]
\label{prop:collusiveParetoOptima}
For an optimal execution strategy problem where two agents minimise costs as in Eq. \eqref{eq:mean_var} in a market as in Eq.~\eqref{eq:perm}, a collusive selling strategy $\Vec{v}^{\,(1,2)}_c$  in the sense of Definition~\ref{defi:collusion} is necessarily a Pareto optimum as defined in Definition~\ref{defi:pareto}. We call this a Collusive Pareto optimal strategy $\Vec{v}^{\,(1,2)}_{cp}$.
\end{prop}
\begin{proof}
By contradiction, suppose that $\Vec{v}^{\,(1,2)}_c$ is not Pareto optimal, then there must exist an iteration $i$ and a strategy $\Vec{u}^{\,(1,2)} = (\vec{u}^{\,(1)}, \vec{u}^{\,(2)})$ such that: 

\begin{equation}
\label{eq:magg}
    \Bar{IS}(\Vec{u}^{\,(1,2)}) \leq \Bar{IS}(\Vec{v}^{\,(1,2)}_c)
\end{equation}
But this contradicts the hypothesis that $\Vec{v}^{\,(1,2)}_c$ is collusive and hence $\Vec{v}^{\,(1,2)}_c$ must be Pareto-optimal.
\end{proof}
Having shown that a collusive strategy is in fact the Pareto-optimal strategy, we aim at finding the set of all Pareto-efficient strategies, i.e. those selling strategies that result in IS levels where it is impossible to improve trading costs for one agent without deteriorating the one of the other agent. In other words, we are looking for those strategies {where an agent is better off only if the other agent is be worse off.} 
We define the set of Pareto solutions and, within this set and we find the Pareto-optimum as the minimum within the set of solutions. 

\begin{defi} [Pareto-efficient set of solutions]
For the two-player game, considering risk neutral players that aim to solve the optimal execution problem in a market model defined as in Eq.~\eqref{eq:perm}, the Pareto-efficient set of strategies is the set of strategies $\Vec{v}^{\,(1,2)}$ such that the IS of one agent cannot be improved without increasing the IS value of the other agent. 
\end{defi}

We now provide the conditions that allow to find the set of Pareto-efficient solutions {as a minimisation of the Implementation Shortfall functional of both the agents. The minimisation has to be intended in the sense of Pareto dominance component-wise.}

\begin{thm}[Pareto-efficient set of solutions]

The Pareto-efficient set of strategies are the solutions to the multi-objective optimisation problem:
    \begin{equation}
    \label{prob:general}
    \tag{P1}
    \begin{cases}
            \min_{\Vec{v}^{\,(1,2)}} \Vec{f}\left(\Vec{v}^{\,(1,2)}\right) \\
            \text{s.t.} \,\, \{ \Vec{v}^{\,(1,2)}\in\mathcal{A},\,\, \Vec{g}(\Vec{v}^{\,(1,2)}) = \Vec{0}\}
    \end{cases}
    \end{equation}
    where:
    \begin{equation*}
    \begin{split}
        \Vec{f}(\Vec{v}^{\,(1,2)}) &= \left[ \Ex\left(IS(\Vec{v}^{\,(1)} | \Vec{v}^{\,(2)})\right) , \Ex\left(IS(\Vec{v}^{\,(2)} | \Vec{v}^{\,(1)})\right) \right]\\
        \Vec{g}(\Vec{v}^{\,(1,2)}) &= \left[ \left(\sum^N_{t=1} v^{(1)}_t - q_0\right)\ , \left(\sum^N_{t=1} v^{(2)}_t - q_0\right)\right]
    \end{split}
    \end{equation*}
\end{thm}
\begin{proof}
Problem~\eqref{prob:general} is a convex multi-objective optimisation problem with $n=N$ design variables $(\Vec{v}^{\,(1,2)})$, $k=2$ objective functions, and $m=2$ constraint functions. Leveraging on \cite{gobbi2015analytical}, in order to find the Pareto-efficient set of solutions, Problem~\eqref{prob:general} can be restated in terms of Fritz-Johns conditions. We define the $\mathbf{L} \in {\mathbb R}^{(n + m) \times (k + m)}$ matrix as:
\begin{equation}
    \mathbf{L} = 
    \begin{bmatrix}
        \nabla_{\Vec{v}^{\,(1,2)}}\Vec{f}(\Vec{v}^{\,(1,2)}) & \nabla_{\Vec{v}^{\,(1,2)}}\Vec{g}(\Vec{v}^{\,(1,2)})\\
        \Vec{0} & \Vec{g}(\Vec{v}^{\,(1,2)})
    \end{bmatrix} 
\end{equation}
then, the strategy $\Vec{v}^{\,(1,2)}_p$ is a solution to the problem:
\begin{equation}
    \label{prob:matrix}
    \tag{P2}
    \mathbf{L} \cdot \Vec{\delta} = \Vec{0}
\end{equation}
where $\Vec{\delta} = (\Vec{\omega}, \Vec{ \lambda}) \in {\mathbb R}^{k+m}$. Moreover, $\Vec{v}^{\,(1,2)}_p$ is a Pareto-efficient solution if $\Vec{\delta}$ exists, it is a non-trivial solution if $\Vec{\delta}$ exists and $\Vec{\delta} \neq \Vec{0}$. 

Leveraging on \cite{gobbi2015analytical}, a solution $\Vec{v}^{\,(1,2), *}$ to the Problem~\eqref{prob:matrix} is a non-trivial Pareto-efficient solution if:
\begin{equation}
\label{eq:det}
    \det(\mathbf{L}(\Vec{v}^{\,(1,2), *})^T\mathbf{L}(\Vec{v}^{\,(1,2), *})) = \Vec{0}
\end{equation}
where $\mathbf{L}(\Vec{v}^{\,(1,2), *})$ denotes the matrix $\mathbf{L}$ where the argument of $f(\cdot)$
and $g(\cdot)$ functions 
is a strategy $\Vec{v}^{\,(1,2), *}$. Thus we say that a necessary condition for generic strategy $(\Vec{v}^{\,(1,2), *})$ to be Pareto-efficient is to satisfy Eq.~\eqref{eq:det}.
In general, Eq.~\eqref{eq:det} gives the analytical formula that describes the Pareto-optimal set of solutions obtainable from Problem~\eqref{prob:matrix}.
\end{proof}

The analytical derivation of the Pareto-efficient set of strategies, i.e. of the Pareto-front, for this kind of problem is quite involved and cumbersome, although obtainable via standard numerical multi-objective techniques. Later in the paper, we will solve this problem numerically to find the set for the problem at hand.  It is instead possible to derive the absolute minimum of the Pareto-efficient set of solutions since the problem is a sum of two convex functions and is symmetric. We call these strategies, one per agent, Pareto-optimal strategies in the spirit of Definition~\ref{defi:pareto}.
\begin{thm}[Pareto optimal strategy]
\label{theo:pareto_strategy}
    For a two-player game with risk-neutral players, the Pareto optimal strategy is the solution of the problem:
    \begin{equation}
            \begin{cases}
                &\argmin_{\Vec{v}^{\,(1)} , \Vec{v}^{\,(2)}} \,\, F(\Vec{v}^{\,(1,2)})\\
                 \text{s.t.}\hspace{1cm} &\left(\sum^N_{t=1} v^{(1)}_t - q_0\right) = 0\\
                &\left(\sum^N_{t=1} v^{(2)}_t - q_0\right) = 0
            \end{cases}
    \end{equation}
    where:
    \begin{equation}
        {F}(\Vec{v}^{\,(1,2)}) = \Ex\left(IS(\Vec{v}^{\,(1)} | \Vec{v}^{\,(2)})\right) + \Ex\left(IS(\Vec{v}^{\,(2)} | \Vec{v}^{\,(1)})\right)
    \end{equation}
     is $\forall\,\,\, t = 1,\dots, N$: 
    \begin{equation}
    \begin{split}
        v^{\,(1)}_{t, p} = \frac{q_0}{N}\,\,, \hspace{2cm} v^{\,(2)}_{t, p} = \frac{q_0}{N}
    \end{split}       
    \end{equation}
    i.e. the TWAP strategy.
\end{thm}
\begin{proof} 
    First define $\Vec{v}^{\,(1)} = \argmin_{\Vec{v}^{\,(1)}} IS (\Vec{v}^{\,(1)} | \Vec{v}^{\,(2)})$, 
    meaning that the strategy for agent $1$ is a function of the strategy of agent $2$. And thus $\Vec{v}^{\,(1)}$ the strategy that minimises the IS of agent $1$ given the strategy of agent $2$. The same consideration holds for agent $2$, where $\Vec{v}^{\,(2)} = \argmin_{\Vec{v}^{\,(2)}} IS (\Vec{v}^{\,(2)} | \Vec{v}^{\,(1)})$.\par
    Leveraging on Proposition~\ref{prop:collusiveParetoOptima}, the Pareto optimal strategies $\Vec{v}^{\,(1)}_p , \Vec{v}^{\,(2)}_p $ of the agents considered solve:
    \begin{equation}
    \label{eq:min_sum}
    \tag{P1}
    \argmin_{\Vec{v}^{\,(1)} , \Vec{v}^{\,(2)}} \,\, F(\Vec{v}^{\,(1,2)})
    \end{equation}
    where:
    \begin{equation}
        {F}(\Vec{v}^{\,(1,2)}) = \Ex\left(IS(\Vec{v}^{\,(1)} | \Vec{v}^{\,(2)})\right) + \Ex\left(IS(\Vec{v}^{\,(2)} | \Vec{v}^{\,(1)})\right)
    \end{equation}
    we then set up a constrained optimisation problem, where the constraint binds the agents to just sell their initial inventory $q_0$ over the time window considered {, where now $\lambda_1, \lambda_2$ are multipliers}:
    \begin{equation}
    \begin{split}
        &\nabla_{\Vec{v}^{\,(1,2)}}\left[{F}\left(\Vec{v}^{\,(1,2)}\right) + \lambda_1\left(\sum^N_{t=1} v^{(1)}_t - q_0\right) + \lambda_2\left(\sum^N_{t=1} v^{(2)}_t - q_0\right) \right]= 0\\
    \end{split}
    \end{equation}
    this can be decoupled into two distinct problems:
    \begin{equation}
        \begin{split}
        &\nabla_{\Vec{v}^{\,(1)}, \lambda_1}  \mathcal{L}_1 = \nabla_{\Vec{v}^{\,(1)}, \lambda_1} \left[ \Ex\left(IS(\Vec{v}^{\,(1)} | \Vec{v}^{\,(2)})\right) + \lambda_1\left(\sum^N_{t=1} v^{(1)}_t - q_0\right)\right] = 0 \\
        &\nabla_{\Vec{v}^{\,(2)}, \lambda_2}  \mathcal{L}_2 = \nabla_{\Vec{v}^{\,(2)}, \lambda_2}  \left[\Ex\left(IS(\Vec{v}^{\,(2)} | \Vec{v}^{\,(1)})\right) + \lambda_2\left(\sum^N_{t=1} v^{(2)}_t - q_0\right)\right] = 0
        \end{split}    
    \end{equation}

    Considering now just the first problem in the previous equation, we notice that:
    \begin{equation*}
        \Ex\left(IS(\Vec{v}^{\,(1)} | \Vec{v}^{\,(2)})\right)=
        -\sum^N_{t=1}\kappa v^{\,(1)}_t\sum^t_{j=1}(v^{\,(1)}_j + v^{\,(2)}_j) - \sum^N_{t=1}\alpha v^{\,(1)^2}_t 
    \end{equation*}
    and thus, for every $t=1, \dots,N$: 
    \begin{equation}
    \begin{cases}
        &\frac{\partial\mathcal{L}_1}{\partial {v}^{\,(1)}_t} = -\kappa \left( (2 q_0 - q^{\,(1)}_t - q^{\,(2)}_t) + {v}^{\,(1)}_t\right)- 2\alpha v^{\,(1)}_t+\lambda_1=0\\
        &\frac{\partial\mathcal{L}_1}{\partial \lambda_1} = \left(\sum^N_{t=1} v^{(1)}_t - q_0\right) = 0
    \end{cases}
    \end{equation}
    where $ \sum^t_{j=1}v^{\,(1)}_j = (q_0 - q_t)$ and $q_t$ is the level of inventory held at time-step $ t = 1, \dots, N$, we notice that:\par
    \begin{equation*}
        v^{\,(1)}_t = -\frac{\kappa(2q_0 - q^{\,(1)}_t - q^{\,(2)}_t) + \lambda_1}{\kappa+2\alpha}
    \end{equation*}
    thus:
    \begin{equation*}
    \begin{split}
        0 &= \sum^N_{t=1}  -\frac{1}{\kappa + 2\alpha}\kappa\left((2q_0 - q^{\,(1)}_t - q^{\,(2)}_t) + \lambda\right) - q_0\\
        \lambda &= \frac{q_0(\kappa + 2\alpha)}{N} + \kappa(2q_0-q^{\,(1)}_t - q^{\,(2)}_t)\\
        v^{\,(1)}_t &= -\frac{\kappa(2q_0 - q^{\,(1)}_t - q^{\,(2)}_t)}{\kappa+2\alpha} + \frac{q_0(\kappa+2\alpha)}{N(\kappa+2\alpha)}+\frac{\kappa(2q_0 - q^{\,(1)}_t - q^{\,(2)}_t)}{\kappa+2\alpha}
    \end{split}
    \end{equation*}
    Finally, the first and the last terms cancel out and we obtain that the Pareto-optimal strategy $\Vec{v}^{\,(1)}_p$ is:
    \begin{equation}
        v^{\,(1)}_{t, p} = \frac{q_0}{N} \,\,\,\,,\,\, \forall\,\,\, t = 1,\dots, N
    \end{equation}
    for $v^{\,(2)}_{t, p}$ same considerations hold since the problem is symmetric.
    
\end{proof}
To study whether \textit{collusive Pareto-optima},  or more generally supra-competitive non-Nash equilibria, arise in this setting, we model the two agents by equally and simultaneously training them with
Double Deep Q Learning (DDQL){, a deep RL algorithm}. 
We then analyse the results under the light of the collusion strategies derived above. The aim is to understand how and which equilibria will eventually be {learned}, and whether in attaining an equilibrium they tacitly collude in order to further drive the costs of their trading down, {sustained by a punishment mechanism}.

{We choose to implement this algorithm and use it for our analysis for two main reasons. On the one hand, using Double Deep Q-Learning allows us to be more consistent with the extant literature that mainly uses Q-learning, outlining if observing tacit-collusive behaviours is still possible when using deep neural networks and memory of past states. In fact, from the literature it is well known that such an off-policy algorithm works well in the single agent case (see, for instance, \cite{macrì2024reinforcement, ning2021double}), making it a safe and sensible ground choice when it comes to extend the framework to more than one agent. On the other hand, and more methodologically, we have chosen Double Deep Q-learning because this algorithm directly maximizes the expected discounted rewards without explicitly parametrizing the policy.
Thus, this algorithm choice allows us to analyse the problem from a purely reward driven point of view, as the Q-values of the agents are the quantities that are optimised through the use of neural networks without the need to visit every state.
We think that an interesting extension of our work could be to compare different RL algorithms to identify pros and cons of each of them.}

\section{Double Deep Q Learning for multi-agent impact trading}
\label{sec:metodi}
We model the algorithmic agents by using RL based on Double Deep Q-Learning (DDQL). The setup is similar to the single agent algorithm as in \cite{ning2021double} and in \cite{macrì2024reinforcement}, but now we consider two agents interacting in the same environment. {The agents are modelled to sequentially interact at each time step to trade either as first or as second. In principle, the game does not need to be a sequential move game but as in real markets, agents cannot trade simultaneously and for this reason we have chosen this numerical setup for the experiments.}
{Moreover, the choice of the algorithm is motivated by the fact that this algorithm has been proven to work with both real data, as in \cite{ning2021double}, and especially with simulated data where the permanent impact and the risk aversion parameter are present as in \cite{macrì2024reinforcement}. 
}

Each agent employs two neural networks, namely the main Q-net ($Q_{\text{main}}$) for action selection and the target Q-net ($Q_{\text{tgt}}$) for state evaluation. The four nets {share the same architecture, their weights evolve independently during the training phase} and they are updated at exactly the same rate. The only difference between the two agents is the timing at which they act and the time at which their nets are updated. In fact, in a given time-step, the agent trading as second pays the price impact generated by the first one. To make the game symmetric, as specified above and similarly to \cite{schied2019market}, at each time step a coin toss decides which agent trades first. This ensures the symmetry and guarantees that, as the game unfolds, no advantage in terms of trading timing is present for either of the two agents.

We divide the overall numerical experiment into  a training and a testing phase. In the training phase we train both the agents to solve the optimal execution problem. In the testing phase, we employ what learnt in terms of $\Qm$ weights, letting the trained agents play an iterated trading game. All the results shown in Section \ref{sec:results} are obtained in the testing phase.

\subsection{Setting of the numerical experiments}
We consider two risk neutral agents whose goal is to unwind an initial position of $q_0=100$ shares with initial value $S_0=10\$$ within a time window $[0, T]$. The window is divided into $N=10$ time-steps of length $T/N=\tau$. The mid-price $S_t$ evolves as in Eq.~\eqref{eq:perm}, and the agents sell their whole inventory during the considered time-window. This is called an \emph{iteration} and in order to train the DDQL algorithm we consider a number of $C$ iterations. This is called a {\it run}. Thus, a run of the game is defined to be a {repeated} game over both $N$ time-steps and in $C$ trading iterations.\par
Over the $C$ iterations, each agent learns how to trade via an exploration-exploitation scheme, thus using $\epsilon$-greedy policies. This is obtained by changing the weights of their $\Qm$ net in order to individually choose the best policies in terms of rewards, related to the obtained IS. The scheme of the algorithm is thus symmetric, and for each agent it is divided into two `phases': an \emph{exploration} and an \emph{exploitation} phase managed by the parameter $\epsilon \in (0,1]$, that is common to both the agents\footnote{{More practically, the agents will explore with probability $\epsilon$ and exploit with probability $1-\epsilon$.}}, that will decrease geometrically during training and globally initialised as $\epsilon=1$. Depending on the phase, the way in which the quantities to sell $v_t$ are chosen changes, and it does so for each agent symmetrically.  When the agents are exploring, they will randomly select quantities to sell $v_t$ in order to explore different states and rewards in the environment, alternatively, when the agents are exploiting, they will use their $Q_{\text{main}}$ net to select $v_t$.
\subsubsection{Action selection and reward function}
For each of the $N$ time-steps and $C$ trading iterations, the agents' knowledge of the state  of market environment is the tuple $g^{i}_t = (t, q^{i}_t, S_{t-1})$ for $i=1,2$. Thus, each agent knows the current time-step, her individual remaining inventory, and the permanently impacted mid-price at the previous time step.
{In our setting \(t \in \{1, \ldots, N\}\) is the discrete time index, \(q_t \in [0, q_0]\) is the agent's remaining inventory at time \(t\), \(S_{t-1} \in \mathbb{R}_+\) is the (permanently impacted) mid-price at the previous time step. All three components are normalised to lie in \([ -1, 1 ]\) before being used as input to the neural network.}

Clearly, only the first and the last are common knowledge of the two agents, while no information on inventory $q_t$ or past selling actions $v_t$ is shared between the agents.\par
Then, depending on the current value of $\epsilon$, a draw $\zeta$, different for each agent, from a uniform distribution determines whether the agent performs exploration or exploitation. This means that with probability $\epsilon$ the agent chooses to explore and thus she chooses at time-step $t$ the quantity to sell $v_t$ sampling from a normal distribution with mean $\mu = \frac{q_t}{N-t}$ and standard deviation $\delta = |\frac{q_t}{N-t}|$. 
Alternatively, with probability $1-\epsilon$, the agent chooses the optimal Q-action as the one that maximises the Q-values from $\Qm$ net, thus exploiting what learnt in the exploration phase. We bind the agents to sell all their inventory within the considered time window, still exploring a large number of states, rewards, and actions.  
\par
Once every $m$ actions taken during training iterations by both the agents, $\epsilon$ is multiplied by a constant $c<1$ such that $\epsilon \to 0$ as $m \to \infty$. In this way, for a large number of iterations $C$, $\epsilon$ converges to zero and the algorithm gradually stops exploring and starts to greedily exploit what the agent has learned in terms of weights $\theta$ of the $\Qm$ net.
Notice that the update for $\epsilon$ happens at the same rate for both the agents, thus they explore and exploit contemporaneously, while the draw from the uniform distribution $\zeta$ is different for every agent. For each time step $t$  we decide which of the two agents trades first with a coin toss. Once that the ordering of the trades is decided, the action decision rule in the training phase unfolds, {for each agent \(i = 1, 2\)}, as:
\begin{align}
    \begin{split}
        \label{eq:action}
        &\epsilon\in(0,1)\,\,,\,\, \zeta_{{i} }\sim\mathcal{U}(0,1)\\
        &v^{(i)}_{{t} } = 
        \begin{cases}
            \sim \mathcal{N}(\mu = \frac{q_{t}}{N-t}, \delta = |\frac{q_{t}}{N-t}|)\,\,\,\,\,\,\,\,\,\,\,\,\,\,\,\,, \text{if}\,\,\zeta \leq \epsilon\\
            \argmax_{v'\in[{0},q_{0}]}Q_{M}(g^{i}_{t},v'|\theta_{\text{main}}) \,\,\,\,\,\,\,\,, \text{else}
        \end{cases}
    \end{split}
\end{align}

{The actions chosen following this rule lie on the strategy space for each agent, which is the set of functions:
    \[
    \pi: (t, q_t, S_{t-1}) \mapsto v_t,
    \]
mapping the agent’s state to a trading action. These policies, denoted by $\pi$, are implicitly encoded by the learnt Q-function, since the agent chooses at each state \(g_t\) the action \(v_t = \arg\max_{v} Q(g_t, v; \theta)\). Thus, the strategy space corresponds to the function class that the Q-network can represent and optimise over. Notice that we do not account for the strategy space directly; indeed the Double Deep Q learning algorithm that we developed and used is an \textit{off-policy} algorithm, thus we do not parametrise the policies but we aim at finding those actions that maximise the Q value which is the output of the Deep Neural Nets (the Q networks), one for each agent.
Notice also that the \textit{action space} for each agent \(i\) at each time step \(t\) consists of trading volumes set to be $v^{(i)}_t \in [0, q^{(i)}_0]$
Actions are also normalised during training. 
}

After this, each agent calculates the reward as:
\begin{equation}
\label{eq:reward}
    r_{t, i} = S_{t-1} v_{t, i} - \alpha v^2_{t, i}\,\,\,\,\,, i=1,2
\end{equation} 
Notice that the actions of the other agent indirectly impact on the reward of the agent $i$ through the price $S_{t-1}$, while nothing but the agent's own actions are known in the reward.\par
Overall, the rewards for every $t\in [1,N]$ are:
\begin{align}
\begin{split}
    \label{eq:rew_expanded}
    r_{1, i} &= S_{0} v_{1, i} - \alpha v_{1, i}^2\\
    r_{2, i} &= (S_1 + \kappa  (v_{1, i} + v_{1, i^-}) + \xi)v_{2, i} - \alpha  v_{2, i}^2\\
    \vdots&\\
    r_{N, i} &= (S_{N-1} + \kappa ( v_{N-1, i} + v_{N-1, i^-})  + \xi)v_{N, i} - \alpha  v_{N, i}^2\\
    \end{split}
\end{align}
Where by $i^-$ we denote the the other agent, assuming that we are looking at the reward for agent $i$, and $\xi \sim \mathcal{N}(0,\sigma)$.\par
Thus, for each time step $t$ the agent $i$ sees the reward from selling $v_{t, i}$ shares, and stores the state of the environment, $g_{t, i}$, where the sell action was chosen along with the reward and the next state $g_{t+1, i}$ where the environment evolves. At the end of the episode, the reward per episode per agent
 is $-q_0 S_0 + \sum^N_{t=1} S_{t-1}v_{t, i} - \alpha v_{t, i}^2$. Written in this form, the aim of the agent is to cumulatively \emph{maximise} such reward, so that the liquidation value of the inventory occurs as close as possible to the initial value of the portfolio.

\subsubsection{Training scheme}

\begin{table}[ht]
  \centering
  \caption{Fixed parameters used in the DDQL algorithm. The parameters not shown in the table change depending on the experiments and are reported accordingly.}
  \label{tab:pmts}
  \begin{tabularx}{\textwidth}{|X|X|X|X||X|X|}
    \hline
    \multicolumn{4}{|l||}{\textbf{DDQL parameters}} & \multicolumn{2}{l|}{\textbf{Model parameters}} \\
    \hline
    NN layers & $5$               & $C$ train its.  & $5,000$    & $N$ intervals     & $10$ \\
    Hidden nodes & $30$           & $M$ test its.   & $2,500$    &  $S_0$ price       & $10\$$\\
    ADAM lr & $0.0001$            & $m$ reset rate & $75$ acts.  &  $q_{0}$ inventory   & $100$\\
    Batch size $b$ & $64$         & $c$ decay rate & $0.995$     &  $\alpha$ t. impact &  $0.002$ \\
    $L$ mem. len. & $15,000$ & $\gamma$ discount & $1$  &   $\kappa$ p. impact               &  $0.001$   \\
    \hline
  \end{tabularx}
\end{table}

Bearing in mind that the procedure is exactly the same for both agents, we focus now on the training scheme of one of them, dropping the subscript $i$. We let the states $g_t$, actions $v_t$, rewards $r_t$ and subsequent future states $g_{t+1}$ obtained by selling a quantity $v_t$ in state $g_t$, to form a \emph{transition} tuple that is stored into a memory of maximum length $L$, we have two different memories, one for each agent. As soon as the memory contains at least $b$ transitions, the algorithm starts to train the Q-nets. To this end, the algorithm samples random batches of length $b$ from the memory of the individual agent, and for each sampled transition $j$ they individually calculate:
\begin{align}
\label{eq:y_T}
y^j_{t}(\theta_{\text{tgt}}) = 
                    \begin{cases}
                        r^j_{t} & \text{if } {t}=N;\\
                        r^j_{t} + \gamma Q_{\text{tgt}}(g^j_{t+1},v^*|\theta_{\text{tgt}}) & \text{else}\quad .
                    \end{cases}
\end{align}
In Eq.~\eqref{eq:y_T}, $r^j_t$ is the reward for the time step considered in the transition $j$, $g^j_{t+1}$ is the subsequent state reached in $t+1$, known since it is stored in the same transition $j$, while $v^* = \argmax_{v}Q_{\text{main}}(g^j_{t},v|\theta_{\text{main}})$. In this context $\gamma$ is a discount factor.  
Each agent individually minimises the mean squared error loss between the target $y(\theta_{\text{tgt}})$ and the values obtained via the $\Qm$ net. In formulae: 
$$    {L}(\theta_{\text{main}},\theta_{\text{tgt}}) = 
                    \frac{1}{b} \sum^b_{\ell =1}\left( \left[  y^\ell_t(\theta_{\text{tgt}}) - 
                    Q_{\text{main}}(g^\ell_t,v^\ell_{t}|\theta_{\text{main}}) \right]\right) ^2$$
$$\theta^*_{\text{main}} = \argmin_{\theta_{\text{main}} }\mathcal{L}(\theta_{\text{main}},\theta_{\text{tgt}}) $$
We then use back propagation and gradient descent in order to update the weights of the $\Qm$ net. This procedure is repeated for each agent and for each random batch of transition sampled from the agent's memory. Overall, once both agents have individually performed $m$ actions,  we decrease $\epsilon$ by a factor $c<1$ and we set $Q_{\text{tgt}}= Q_{\text{main}}$\par
Once both agents have been simultaneously trained to optimally execute a quantity $q_0$ while interacting only through the midprice, we let the agents interact on another number $M<C$ of trading iterations. 
This is the \emph{testing phase}, and now actions for each agent are selected using just her $Q_{\text{main}}$ net. As said above, the results analysed below are those obtained in the testing phase.\par

The features of the Q-nets for each agent are ($q_{i,t},t, S_{t-1},v_{i,t}$) and are normalised in the domain $[-1,1]$ using the procedure suggested in \cite{ning2021double}, whereas normalised mid-prices $\Bar{S}_t$ are obtained via min-max normalisation.

{When considering the inputs to the Q-networks, we choose this set of features for two main reasons. The first one is more theoretical. We want to model the interaction between the agents in such a way that there is no direct exchange of information about the strategy followed by either of them. With this choice of features, the only available `shared' information is the one on the the mid-price level of the traded security, thus no direct communication between the two agents happens. The second reason is more practical. On the one hand we want to be parsimonious with the set of features seen by the agent, and since the environment where the agents trade is a multi-agent Almgren and Chriss model, the inventory and the time step along with the traded quantity, would suffice as features since the problem would be solved with strategies that do not depend on the level of mid-price. However, as shown in \cite{ning2021double, macrì2024reinforcement} adding additional features that theoretically should be irrelevant, practically help the algorithm learning the optimal trading strategy faster.}

In our setup we use fully connected feed-forward neural networks with $5$ layers, each with $30$ hidden nodes. The 
activation functions are leakyReLu, and finally we use ADAM for optimisation. With the exception of the volatility parameter $\sigma$ which will be specified later, the parameters used in the algorithm are reported in Table~\ref{tab:pmts}, the training algorithm is reported in Algorithm~\ref{alg:cap}.

\FloatBarrier
\begin{algorithm}[ht]
\caption{Training of Double Deep Q-Learning multi-agent impact game}
\begin{algorithmic}
\Require
{\\
Set $\epsilon = 1$, $b$ batch size, $C$ train iterations;\\
Set market dynamics parameters;\\
For each agent $i = 1,2$ initialise with random weights $Q_{\text{main}}$ and make a copy $Q_{\text{tgt}}$ ;\\
For each agent $i = 1,2$ initialise the memory with max length $L$. \\
}
\For {k in C}
    \State Set $S^k_0 = S_0$;\\
    \For{t in N}
        \State{$u \sim \text{Bin}(1, 0.5)$; } \Comment{Decides the order of execution}
        \State{\If{$u = 0$} $i = (1, 2) \,\,\,\text{else}\,\,\, i = (2, 1)$\EndIf} \Comment{Vector of priority ordering of the agents}
        \For{each agent $i$ in the order decided above}
            \State {$g_{i,t} \gets (q_{i,t},t, S^k_t)$;}  
            \State {$v_{i, t} \gets          \begin{cases}
            \text{sample}\,\,\, \mathcal{N}(\mu = \frac{q_{t}}{N-t}, \delta = |\frac{q_{t}}{N-t}|)\,\,\,\,\,\,\,\,\,\,\,\,\,\,\,\,, \text{with probability $\epsilon$}\\
            \argmax_{v'\in[0,q_{0}]}Q_{M}(g_{t},v'|\theta_{\text{main}}) \,\,\,\,\,\,\,\,, \text{with probability ($1-\epsilon$)}
        \end{cases};$}\\
            \State{$r_t\gets S^i_{t-1} v_t - \alpha v^2_t$;}\\
            \State{$S^k_{t-1}\to S^k_t$;}\Comment{Generate $S^k_t$ from $S^k_{t-1}$}\\
            \State{$g_{i, t+1} \gets (q_{i, t+1}, t+1, S^k_{t})$ ;}\\
            \State {Memory for agent $i$ $\gets(g_{i,t},r_{i,t},v_{i,t}, g_{i,t+1})$;} \Comment{Memory storing}
            \\
            \If{Length of memory $\ge$ $b$}
                \For{j in $b$}
                    \State{Sample a batch of $(g^j_{i,t},r^j_{i,t},v^j_{i,t},g^j_{i,t+1})$ from memory;}
                    \State{$v^*_{i} = \argmax_{v}Q_{i, \text{main}}(g^j_{i,t},v^j_{i, t}|\theta_{i,\text{main}})$;}
                    \State{$y^j_{i,t}(\theta_{i,\text{tgt}}) = 
                        \begin{cases}
                            r^j_{i,t} & \text{if } {t}=N;\\
                            r^j_{i,t} + \gamma Q_{i,\text{tgt}}(g^j_{i,t+1},v^*_{i}|\theta_{i,\text{tgt}}) & \text{else}
                        \end{cases}$} 
                \EndFor\\
                \State{$\theta^*_{i,\text{main}} = \argmin_{\theta_{i,\text{main}}}\mathcal{L}(\theta_{i,\text{main}},    \theta_{i,\text{tgt}})$ via gradient descent with loss to minimise:
                        $${L}(\theta_{i,\text{main}},\theta_{i,\text{tgt}}) = 
                        \frac{1}{b} \sum^b_{\ell =1}\left( \left[  y^\ell_t(\theta_{i,\text{tgt}}) - 
                        Q_{i,\text{main}}(g^\ell_{i,t},v^\ell_{i,t}|\theta_{i,\text{main}}) \right]\right) ^2$$}
                \If{Length of agent $i$ memory = $L$} halve the length of agent $i$ memory\EndIf
            \EndIf
            \State{After $m$ iterations decay $\epsilon=\epsilon\times c$;}
            \State{After $m$ iterations $\theta_{i,\text{tgt}}\gets\theta_{i,\text{main}}$;}
        \EndFor
    \EndFor
\EndFor
\end{algorithmic}
\label{alg:cap}
\end{algorithm}
\FloatBarrier

\begin{figure}[ht]
    \centering
    \includegraphics[scale=0.6]{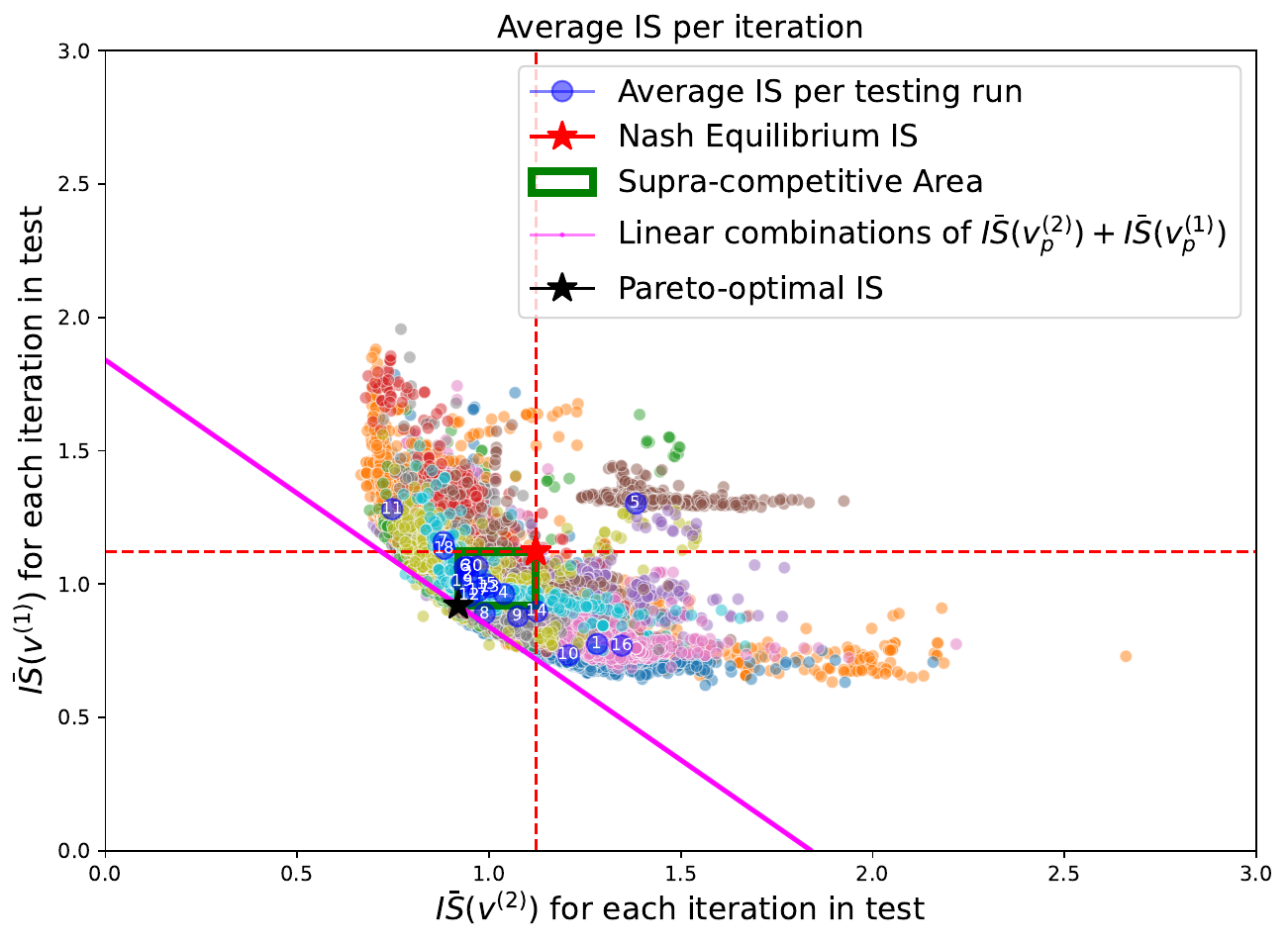}
    \caption{Scatter plot of the IS of the two agents for 20 testing runs of $2,500$ iterations in the zero noise case ($\sigma= 10^{-9}$).}
    \label{fig:scatter_0}
\end{figure}

\section{Results}\label{sec:results}

In this Section we present the results obtained by using the RL algorithm {shown in Algorithm~\ref{alg:cap}} in the game setting outlined in Section~\ref{sec:mkt_set}. The experiments aim at studying the existence and the form of the {equilibrium strategies and of their costs, learnt by the agents during the experiments, this is done} 
by analysing the policies chosen by the two interacting trading agents. We then compare the equilibria {under the learnt dynamics} with the Nash equilibrium strategies, the Pareto-efficient set of solutions, and the Pareto optimal strategy. We consider different scenarios, using the market structure as outlined in Eq.~\eqref{eq:perm}. 

It can be easily noticed how the theoretical Nash equilibrium for this game does not depend on the volatility level of the asset. In fact, when the risk aversion parameter $\lambda = 0$, the Nash equilibrium of Eq.~\eqref{eq:opt_inv_two_players} depends only on the permanent and temporary impact coefficients $\kappa$ and $ \alpha$, respectively. However, in the numerical determination of the equilibrium solution, volatility $\sigma$ and the associated diffusion play the role of a disturbance term, since an agent cannot determine whether an observed price change is only due to the impact generated by the trading of the other agent or if it is a random fluctuation due to volatility. In this sense, volatility plays the role of a \textit{noise-to-signal} ratio here. 
{One can interpret the level of \(\sigma\) as a measure of presence of noise traders that are randomly affecting the price while the two agents do optimal execution. In this sense, we could also consider \(\sigma\) as a measure of market liquidity.}
To quantify the effect of volatility on the learnt equilibria, we perform three sets of experiments for different levels of the volatility parameter $\sigma$, leaving unchanged the impact parameters. More specifically we consider, for both training and testing phase the case where  $\sigma = 10^{-9}$ (termed \textit{zero noise} case), $\sigma = 10^{-3}$ (\textit{moderate noise}) and  $\sigma=10^{-2}$ (\textit{large noise}) case. In all three cases, we use the same temporary and permanent impacts $\alpha=0.002$ and $\kappa=0.001$. 

{Other preliminary analyses on the behaviour of the agents when one agent follows a Nash or a Pareto strategy, and the other learns a reaction strategy can be found in Appendix~\ref{app:other_experiments}. In what follows we will only concentrate on the simultaneous multi-agent learning problem.}

We employ 20 training and testing runs, each run is independent from the others, meaning that the weights found in one run are not used in the others. In this way we aim at independently (run-wise) train the agents to sell their inventory. Each testing run has $M = 2,500$ iterations of $N = 10$ time-steps.

\subsection{The \emph{zero noise} case}
When $\sigma = 10^{-9}$ we model the interaction of both agents in a limiting situation when the market is very illiquid and almost no noise traders enter the price formation process. 
In this case, the permanent impact is essentially the only driver of the mid-price dynamics in Eq.~\eqref{eq:perm}, thus price changes are triggered basically only by the selling strategies used by the agents throughout the game iterations.

The results of the experiments are displayed in Figure~\ref{fig:scatter_0}, which shows, as a scatter plot, the $IS$ of both agents in each iteration of the testing phase. Each colour represents the results of one of the $20$ runs of $2,500$ iterations each. The blue circles show the $IS$ centroids per testing run (the number in the circle identifies the run). For comparison the plot reports as a red star the point corresponding to the Nash equilibrium of Eq.~\eqref{eq:opt_inv_two_players}, which is used to divide the graph into four quadrants, delimited by the red dashed lines. The top-right quadrant contains point which are sub-optimal with respect to the Nash equilibrium for both players, while the points in the top-left and bottom-right quadrant report points where the found equilibrium favours one of the two agents at the expenses of the other. In particular, in these regions one of the agents achieve an IS smaller than the one of the Nash equilibrium, while the other agents performs worst. In a sense, one of the agents predates on the other in terms of reward. The bottom-left quadrant is the most interesting, since here {\it both} agents are able to achieve an $IS$ smaller than the one in the Nash equilibrium and thus contain {supra-competitive \(IS\) costs that are} potential collusive equilibria. For comparison, the black star indicates the $IS$ of the Pareto-optimal strategy found in Theorem~\ref{theo:pareto_strategy}, while the magenta line is the linear combination of the Pareto-optimal $IS$. Finally, the green rectangle denotes the area between the Nash and the Pareto-optimal $IS$s, and we call it the \textit{supra-competitive area}. In fact, in that area we find the $IS$s for both the agents that lie between the \textit{ collusion} defined by the Pareto-optimum and the Nash equilibrium. {It is the area where, although the $IS$ costs achieved do not overlap perfectly with the Pareto optimal one, thus the collusion, are still supra-competitive}.

\begin{figure}[t]
    \centering
    \includegraphics[scale=0.45]{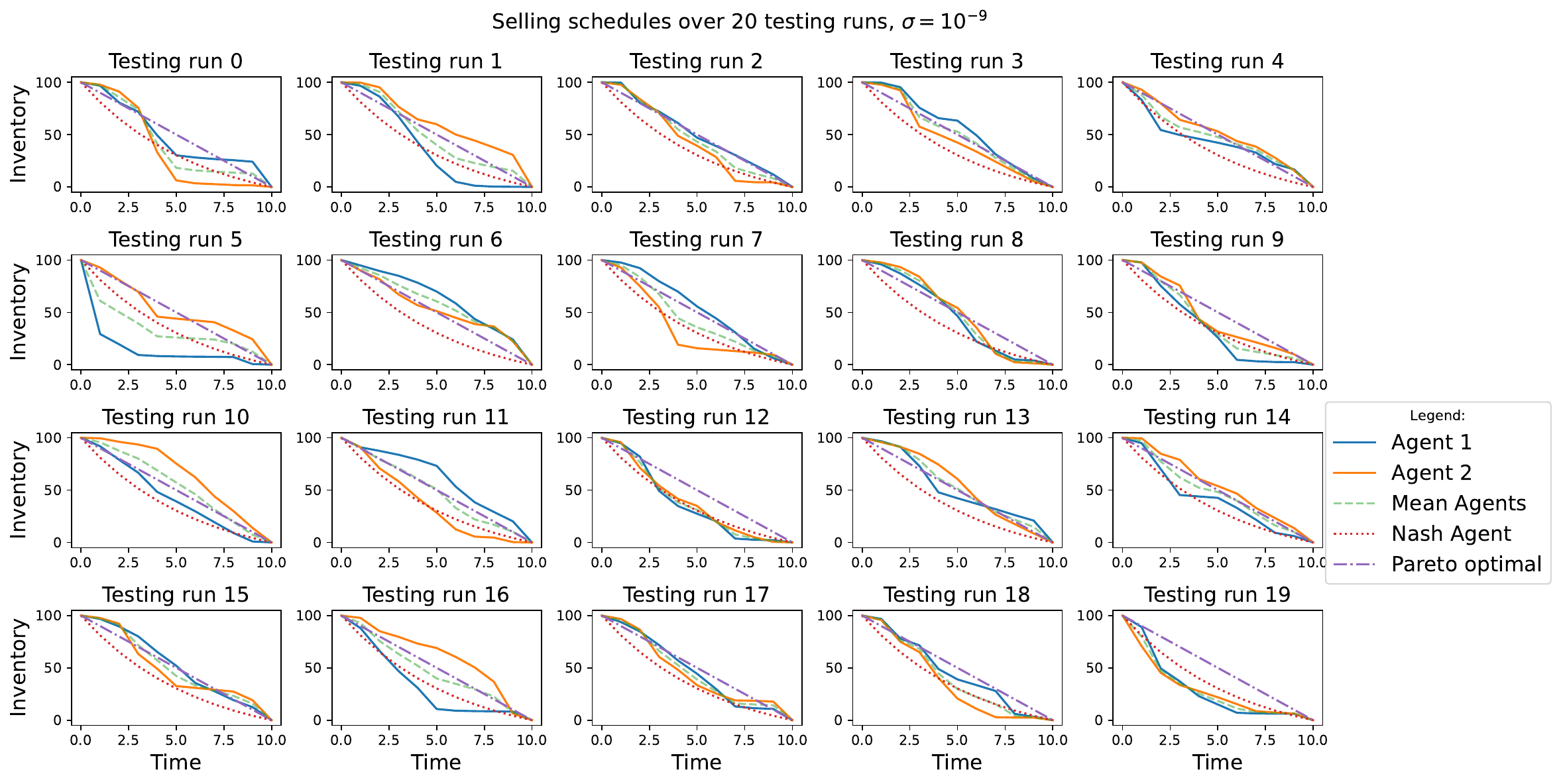}
    \caption{Optimal execution strategies for 20 testing runs of $2,500$ iterations, using $\sigma= 10^{-9}$.}
    \label{fig:trading_sigma0}
\end{figure}

Looking at Figure~\ref{fig:scatter_0} we can notice that the $IS$s per testing run concentrate in the {supra-competitive} area of the graph (green rectangle) between the Nash and the Pareto-optimal $IS$. This means that when the noise is minimal, it is easier for the agents to adopt {strategies whose cost levels are supra-competitive and potentially consistent with }tacit collusive strategies. The agents are able to obtain costs that fall near the collusion $IS$, i.e. the Pareto-optimum. The majority of the remaining $IS$s per iteration still lie within the second and fourth quadrant of the graph, meaning that again they are in general able to find strategies that, at each iteration, do not allow one agent to be better without worsening the other.

Considering the strategies found by the agents in Figure~\ref{fig:trading_sigma0}, we notice how the agents trade at different speeds, i.e. their selling policies are consistent with the presence of a slow trader and a fast trader. It can be seen how the policies followed by the agents in almost all the 20 simulations depend inversely on the strategy adopted by the competitor. 
Moreover, in most of the runs in the {supra-competitive} region, the \textit{average} strategy of the agents is very similar to the TWAP strategy. This is possible thanks to very low noise, and in this case the agents are able to find more easily the Pareto-optimal strategy, which corresponds to an equilibrium where they both pay the lowest amount possible in terms of $IS$. 

These evidences underline the quite intuitive fact that, when the noise-to-signal ratio is low it is simple for the agents to disentangle their actions from those of other agents, {thus making it easier for the agents to learn and possibly infer the strategy of the competitor}. The agents find an incentive to deviate from the Nash-equilibrium adopting strategies that allow for lower costs. The majority of these strategies correspond to an average {supra-competitive} $IS$, thus with lower cost than in the Nash equilibrium, but still slightly greater than the Pareto-optimum in terms of $IS$. In general, per iteration neither agent can be better off without increasing the other agent cost level. 

Coupling the cost plot in Figure~\ref{fig:scatter_0} with the average selling schedules in Figure~\ref{fig:trading_sigma0}, it can be seen how the centroids that relate to higher costs for one of the agents correspond to a deviation from the TWAP strategy by either of them, {such deviation corresponds to another deviation by the other agent, that substantiates in a faster selling rate. There is, on the one hand, evidence of supra-competitive behaviour, on average, when $IS$ costs are considered}, which in turn is reflected in the way the agents trade at different speeds. The phenomenon is evident thanks to the very low level of asset's volatility. Thus, {hinting to tacit collusive behaviour in this first case, judging by the supra-competitive cost outcomes. In Section 5 we thoroughly discuss  this point.} 

\begin{figure}[t]
    \centering
    \includegraphics[scale=0.6]{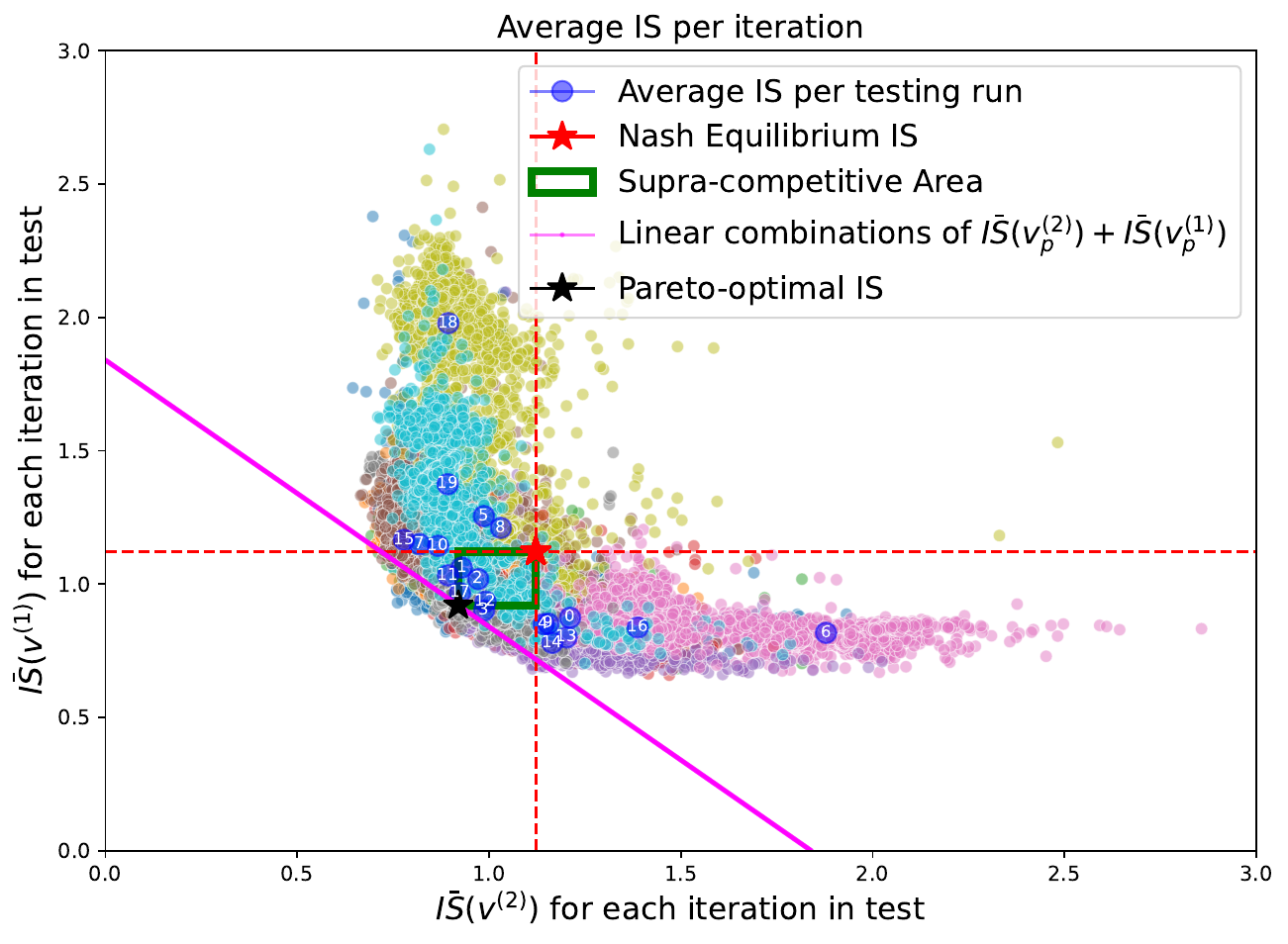}
    \caption{Scatter plot of the IS of the two agents in 20 testing runs of $2,500$ iterations in the moderate noise case ($\sigma= 10^{-3}$).}
    \label{fig:scatter_001}
\end{figure}

\subsection{The \emph{moderate noise}  case}

 In this setting, the mid-price dynamics is influenced both by the trading of the agents and by the volatility. The first thing that can be noticed in Figure~\ref{fig:scatter_001} is that the $IS$ per iteration distributes almost only in the second, third and fourth quadrant of the plot. Moreover, the distribution of the points suggests again an inverse relation between the $IS$s of the two agents, i.e. a lower $IS$ of an agent is typically associated with a worsening of the $IS$ of the other agent. The centroids distribute accordingly. In fact, the centroids that lie in the supra-competitive area of the graph  are as numerous as those that lie outside of that area on either the second or fourth quadrant. It can be seen that, outside the green rectangle, the agents tend to behave in a \textit{predatory} way, meaning that one agent has consistently lower costs than the other. This happens tacitly, meaning that no information about the existence or about the strategy followed by the other agent is part of the information available either in the market or in each agent's memory. Notice that predatory strategies are still consistent with the definition of Pareto efficiency but are \textit{not a proper collusion} since they do differ from the Pareto-optimal $IS$, even if they appear to overlap in Figure~\ref{fig:scatter_001}.
\FloatBarrier
\begin{figure}[ht]
    \centering
    \includegraphics[scale=0.45]{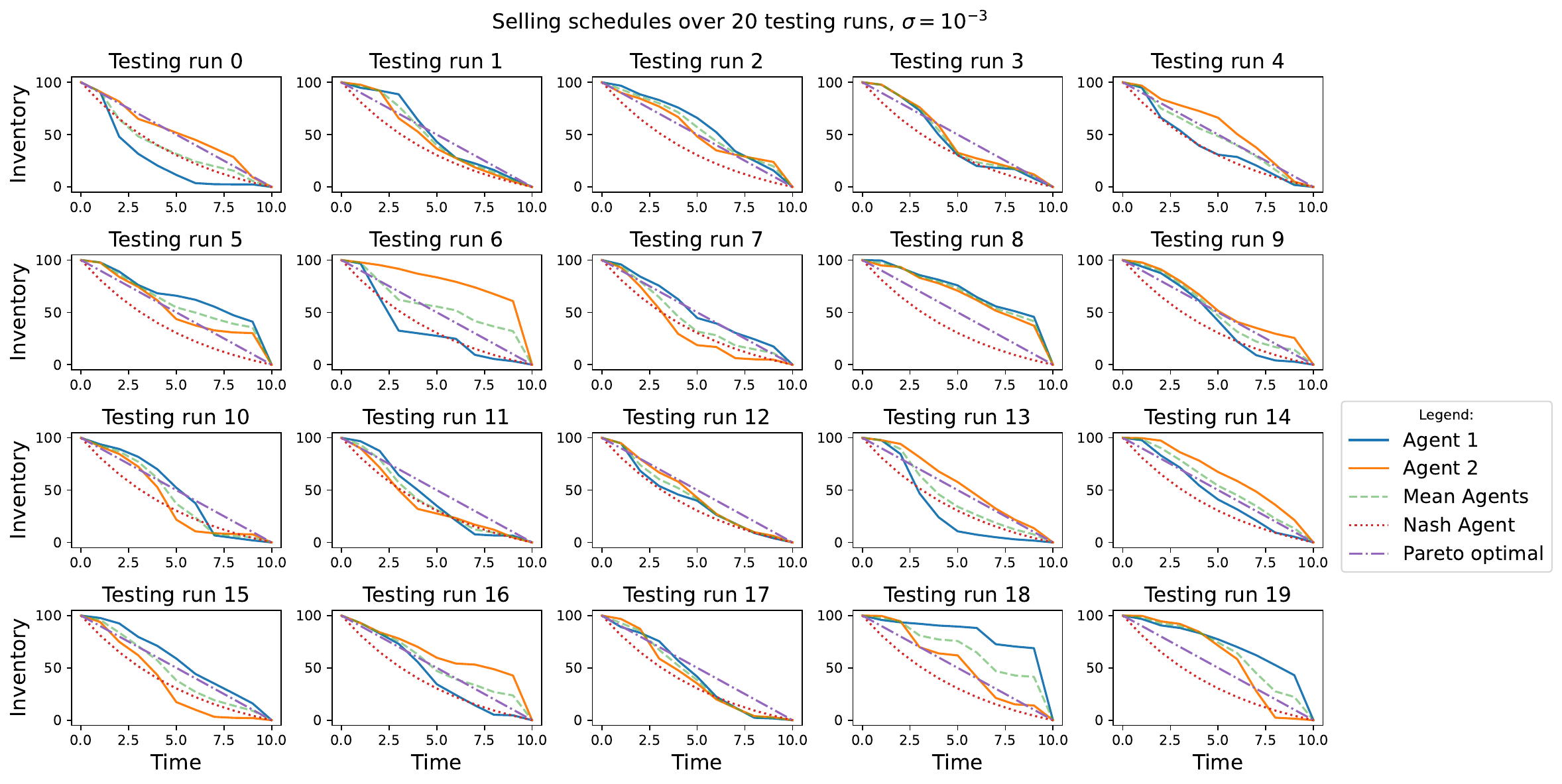}
    \caption{Optimal execution strategies for 20 testing runs of $2,500$ iterations, using $\sigma= 10^{-3}$.}
    \label{fig:trading_sigma001}
\end{figure}
\FloatBarrier
Looking at the average strategies implemented by the agents in each testing run (Figure~\ref{fig:trading_sigma001}), it can be noticed how as in the {\it zero noise} case, roughly speaking there is almost always one agent that trades faster than the other and, again as in the previous case, the one agent that trades slower is the one with a higher $IS$. The comparison of Figure~\ref{fig:trading_sigma001} and Figure~\ref{fig:scatter_001} shows that the agent that `predates' the other is a fast trader and obtains lower costs of execution. This behaviour is more pronounced in this case, and we postulate that this is due to the increased level of noise for this experiment. 

We conclude once again that, even if no explicit information about the trading strategies is shared by the agents, during the repeated game they are able to extrapolate information on the policy pursued by the competitor through the changes in the mid-price triggered by their own trading and by the other agent's trading. Thus, it seems plausible that RL-agents modelled in this way are able to extrapolate information on the competitor's policy and tacitly interact through either a {plausibly cooperative} behaviour that leads to supra-competitive costs or through a predatory behaviour where the costs of one agent are consistently higher than those of the other. Generally speaking, the set of solutions achieved is in line with the definition of the Pareto efficient set of solutions.

\subsection{The \emph{large noise} case}
\begin{figure}[ht]
    \centering
    \includegraphics[scale=0.6]{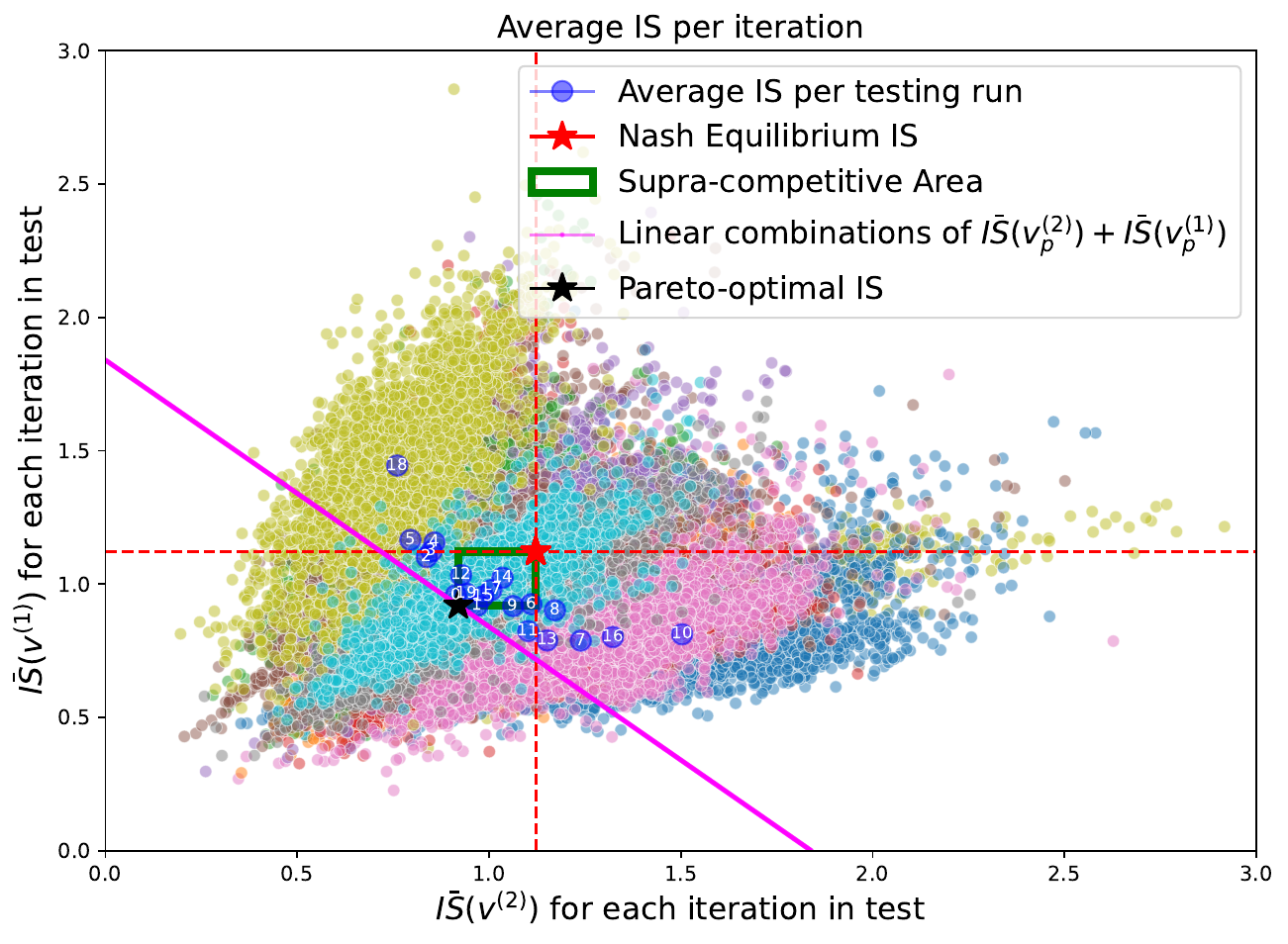}
    \caption{Scatter plot of the IS of the two agents for 20 testing runs of $2,500$ iterations in the large noise case ($\sigma= 10^{-2}$).}
    \label{fig:scatter_01}
\end{figure}

Finally, we study the interactions between the agents when the volatility of the asset is large. This corresponds to a relatively liquid market, since price changes are severely affected by the volatility level {and the price impact triggered by the agents is low in comparison}. Looking at Figure~\ref{fig:scatter_01}, we notice that in this market setup the higher level of volatility significantly affects the distribution of the $IS$ per iteration. In fact, the points in the scatter plot now distribute obliquely, even if for the most part they still lie in the second, third, and fourth quadrants. Because of  the higher volatility level, very low $IS$ values for both agents might be attained per iteration and the structure of the costs per iteration is completely different with respect to to the previous two cases. However, the distribution of the average $IS$s per testing run, i.e. the positions of the centroids, is still concentrated for the greatest part in the strategies' area in between the Pareto-optimum and the Nash equilibrium costs. 

\FloatBarrier
\FloatBarrier

\begin{figure}[ht]
    \centering
    \includegraphics[scale=0.45]{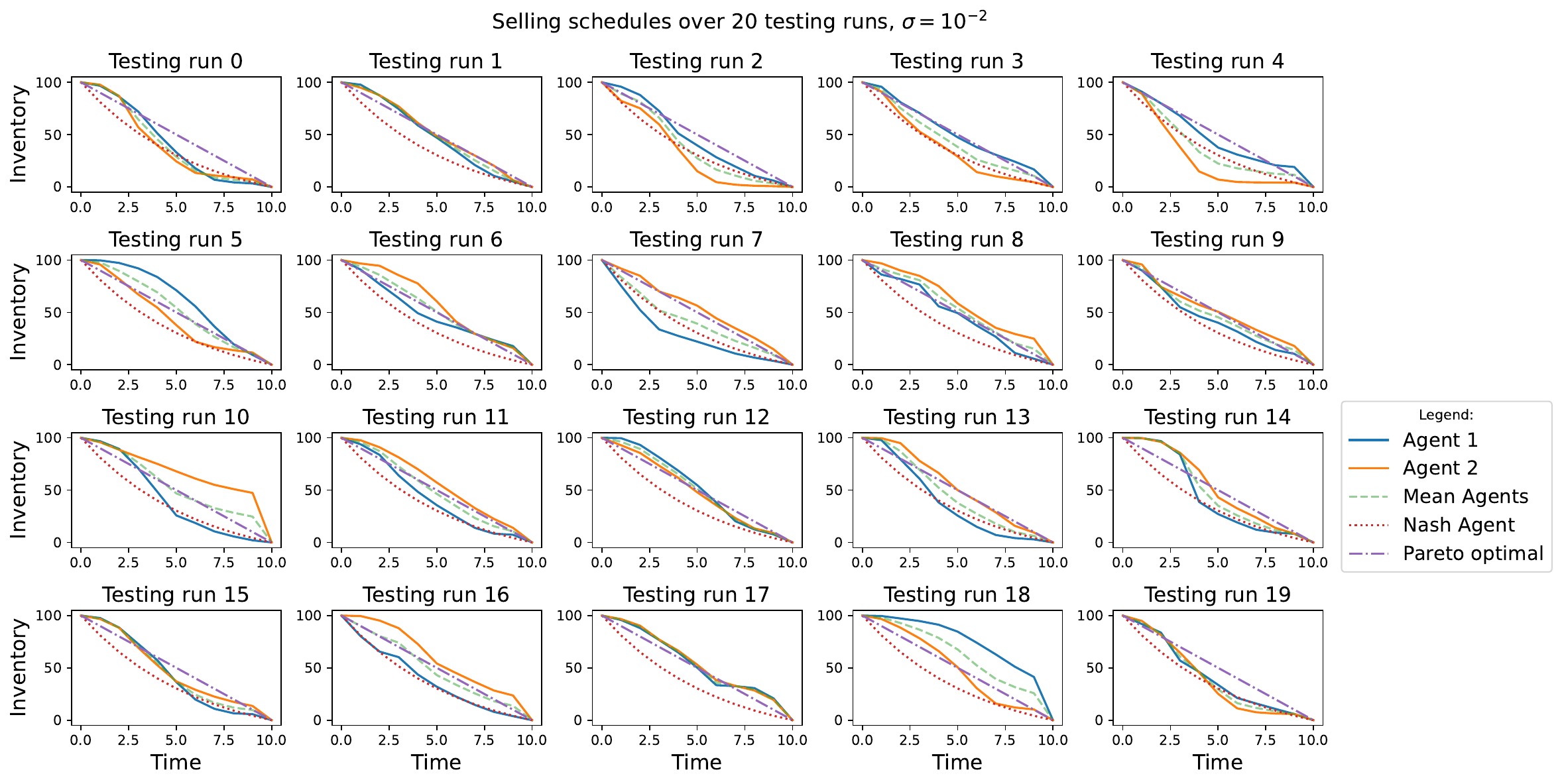}
    \caption{Optimal execution strategies for 20 testing runs of $2,500$ iterations, using $\sigma= 10^{-2}$.}
    \label{fig:trading_sigma01}
\end{figure}

 Figure~\ref{fig:trading_sigma01} shows the trading strategies for the agents. It can still be noticed how in general if one agent is more aggressive with its trading, the other tends to be not. This behaviour of the agents, resulting in faster and slower traders, still takes place and again the agent that is trading slower pays more in terms of $IS$. {Comparing} the centroids' distribution and the selling schedules in Figure~\ref{fig:trading_sigma01} and Figure~\ref{fig:scatter_01} reveals how a slower trading speed worsens the cost profile of one agent to the benefit of the other. The average trading strategy of the agents per testing run in the {supra-competitive} cases basically revolves around a TWAP strategy that is in turn the Pareto-optimum for the problem considered.

\subsection{Summary of results and comparison with the Pareto front}

\begin{figure}[ht]
    \centering
    \includegraphics[scale=0.6]{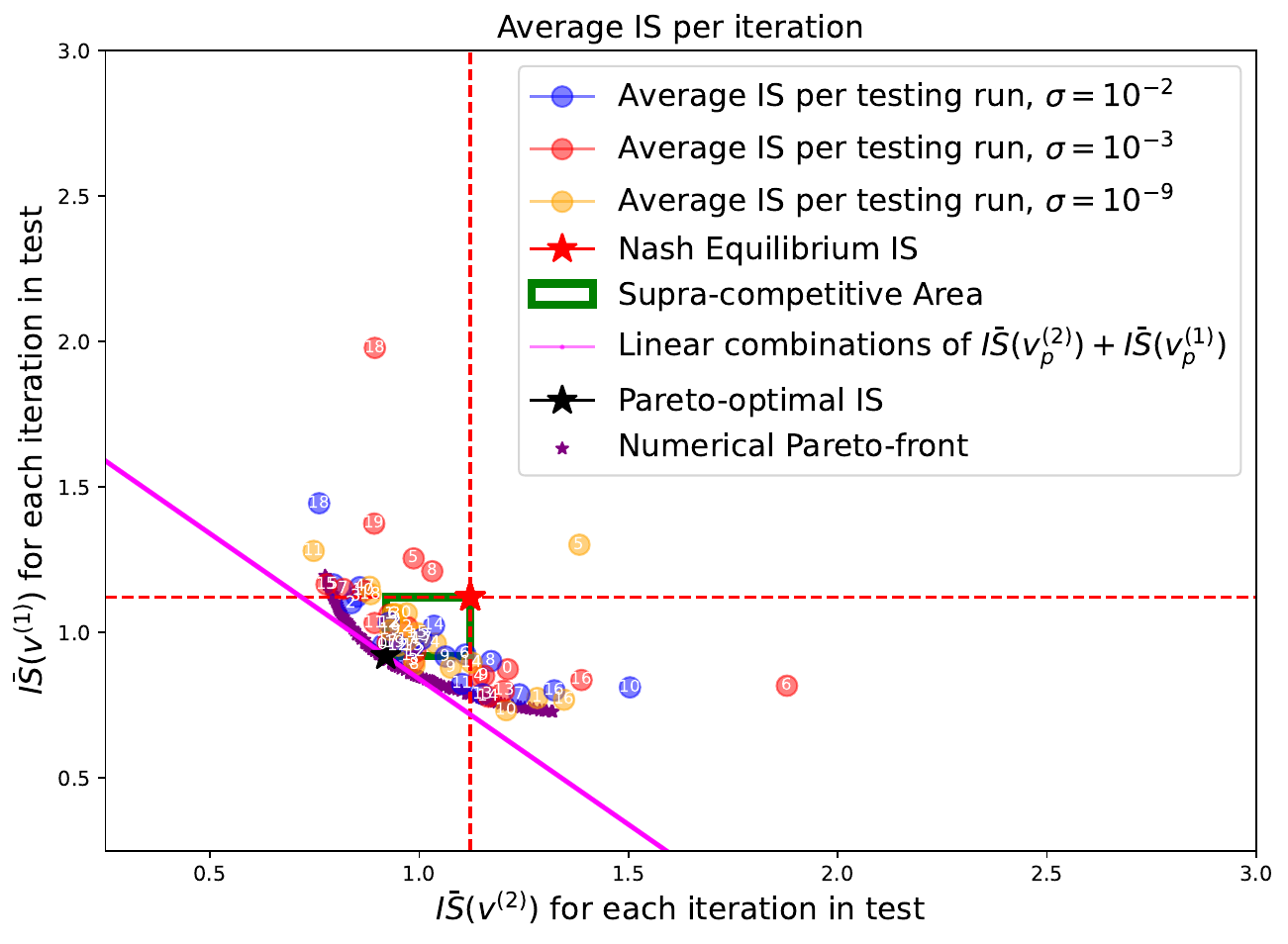}
    \caption{Scatter plot of the IS centroids  of the two agents in the $20\times 3$ testing runs of $2,500$ iterations each, for all the considered values of $\sigma$.}
    \label{fig:scatt_tot}
\end{figure}

We have seen above that especially in the presence of significant volatility, the points corresponding to the iterations in a run tend to distribute quite widely in the scatter plot. The centroid summarises the average behaviour in a given run and provides a much more stable indication of the relation between the $IS$ of the two agents.
To have a complete overview of the observed behaviour across the three volatility regimes, in Figure~\ref{fig:scatt_tot} we show in a scatter plot the position of the centroids of the $20\times 3$ testing runs. It can be seen that the centroids tend to concentrate in the {supra-competitive  area}, irrespective of the volatility level for the experiment. We notice how the agents' costs tend to cluster in this area and to be close to the Pareto-optimal $IS$. To have a more detailed comparison of the simulation results with the theoretical formulation of the game, we numerically compute the Pareto-efficient set of solutions (or Pareto-front)\footnote{The numerical Pareto front has been obtained using `Pymoo' package in Python introduced in \cite{pymoo}.} and represent it on the scatter plot. The Figure shows that, as expected, the centroids lie on the top-right of the front. More interestingly, they are mostly found between the front and the Nash equilibrium  for all the levels of volatility and, in line with the definition of the Pareto-front, for an agent to get better the other has to be worse off. We further notice that the majority of points for the zero noise case lies very close the Pareto optimum or in the supra-competitive area, thus the lower the volatility the easier it is to converge to a supra-competitive strategy. In the other two cases we notice how, even roughly half of the centroids lie in the supra-competitive area, the rest  mostly lie near the numerical Pareto front.
\FloatBarrier
\FloatBarrier
\FloatBarrier
\begin{figure}[ht]
    \centering
    \includegraphics[scale=0.45]{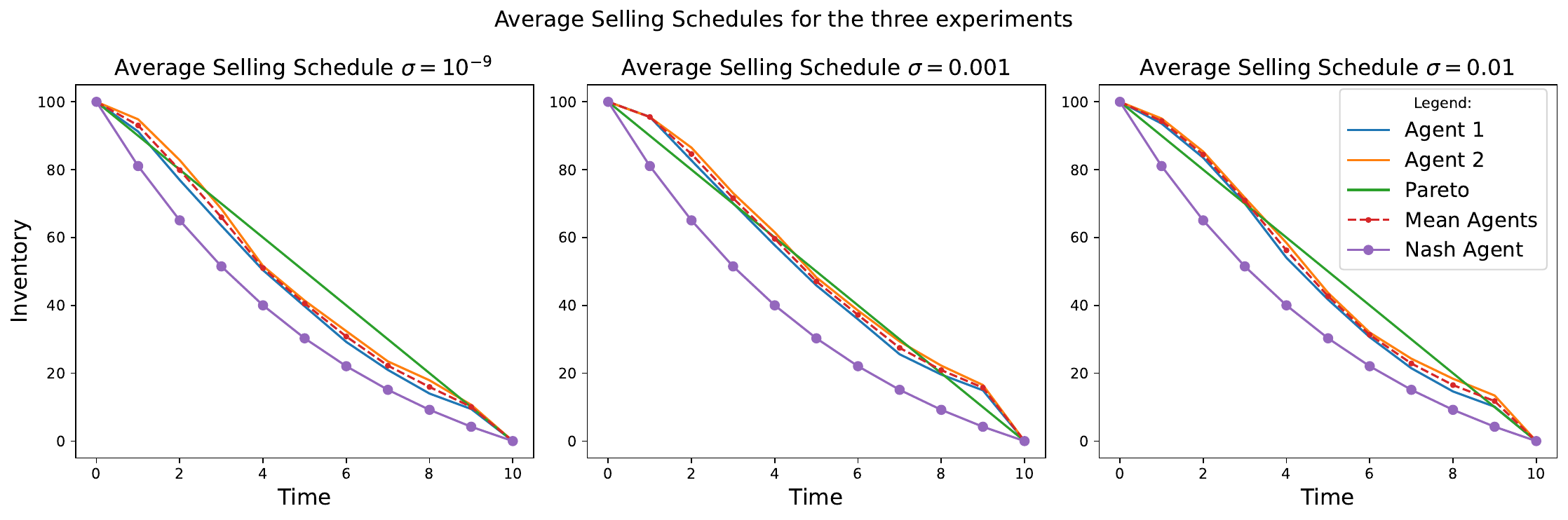}
    \caption{Average optimal execution strategy over the 20 testing runs of $2,500$ iterations, for all the values of $\sigma$ considered.}
    \label{fig:trading_tot}
\end{figure}
\FloatBarrier
 Figure~\ref{fig:trading_tot} shows that, irrespective of the values considered for the volatility level $\sigma$, the average strategy does not differ much from the Pareto-optimum TWAP strategy. In the \textit{zero noise} case, the average strategy of the two agents tends to be slightly more front loaded, {i.e. they tend to liquidate more at the beginning of the trading schedule}, but still lies between the Nash and the Pareto-optimum. As the $\sigma$ value increases, and thus in the \textit{moderate noise} and \textit{large noise} cases, we can see how both the agents tend to be less aggressive at the beginning of their execution in order to adopt a larger selling rate towards the end of their execution. This behaviour is stronger the higher the $\sigma$, due to the increased uncertainty brought by the asset volatility in addition to the price movements triggered by the trading of both the agents.

\subsection{Variable volatility and misspecified dynamics}

Our simulation results show that volatility plays an important role in determining the $IS$ in an {episode} of the testing phase, although when averaging across the many iterations of a run, the results are more stable and consistent. In financial market, returns are known to be heteroscedastic, i.e. volatility varies with time. Thus, also from the practitioner's point of view, it is interesting to study how agents, which are trained in an environment with a given level of volatility level, perform in a testing environment where the level of volatility is different. In particular, it is not a priori clear what behaviour among the two agents would naturally arise when the price dynamics is different in the testing and in the training phase. Moreover, it is interesting to study how a change in the environment would impact the overall selling schedule and the corresponding costs. In the following, we study the agents' behaviour in extreme cases, in order to better appreciate their behaviour under time-varying conditions. Specifically, using the same impact parameters as above in both phases, we learn the the weights of the DDQN algorithm in a training setting  with $\sigma = 10^{-9}$ and then we employ them in a testing environment where $\sigma = 10^{-2}$. We then repeat the experiment in the opposite case with the volatility parameters switched, i.e. training with $\sigma = 10^{-2}$ and testing with $\sigma = 10^{-9}$. We now run $10$ testing runs of $2,500$ iterations each.

\begin{figure}[ht]
    \centering
    \includegraphics[scale=0.6]{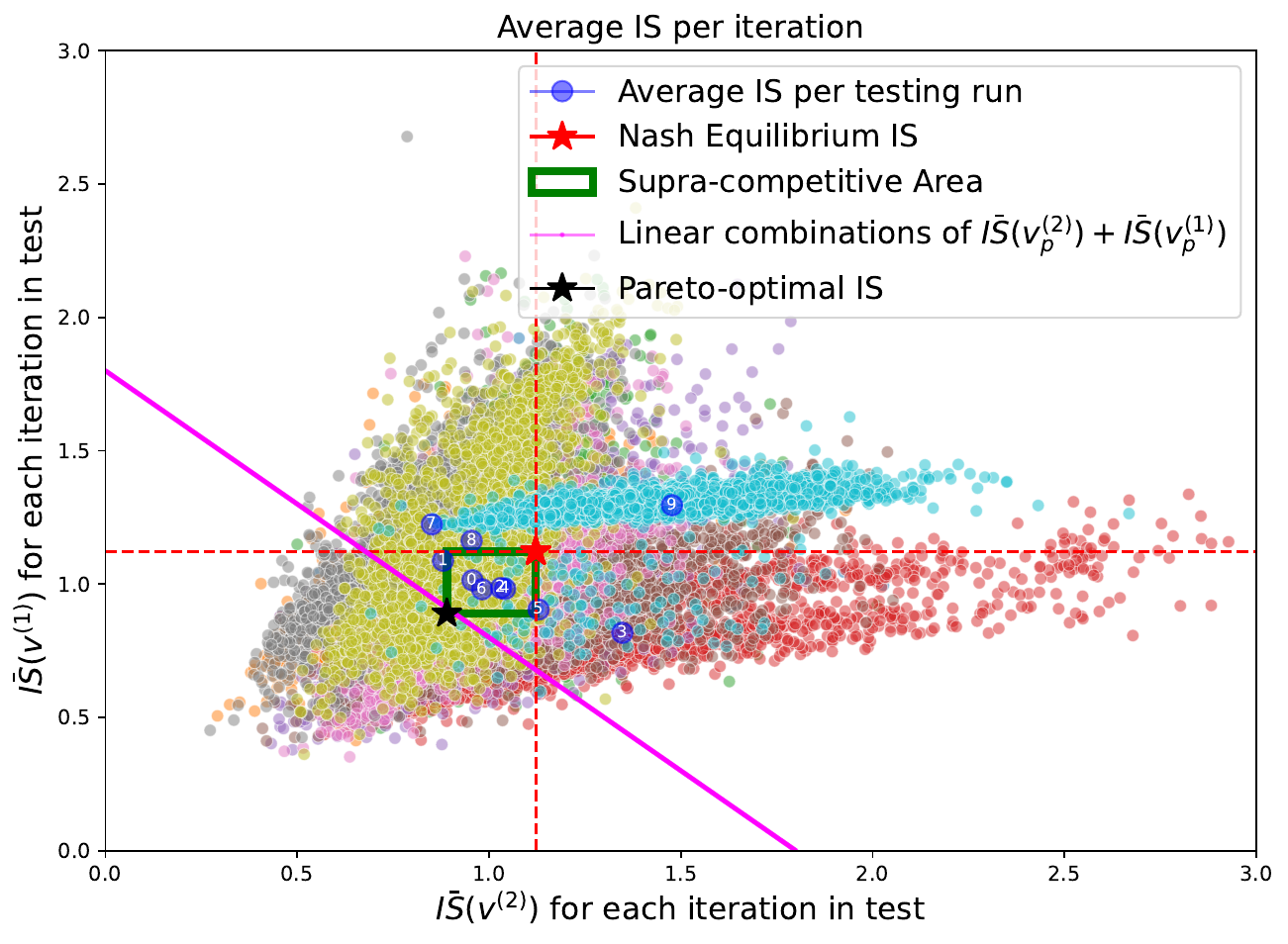}
    \caption{IS scatter for $20$ testing runs of $2,500$ iterations, training $\sigma = 10^{-9}$ and testing $\sigma = 10^{-2}$.}
    \label{fig:scatter_mix_0_test_1}
\end{figure}
\begin{figure}[ht]
    \centering
    \includegraphics[scale=0.4]{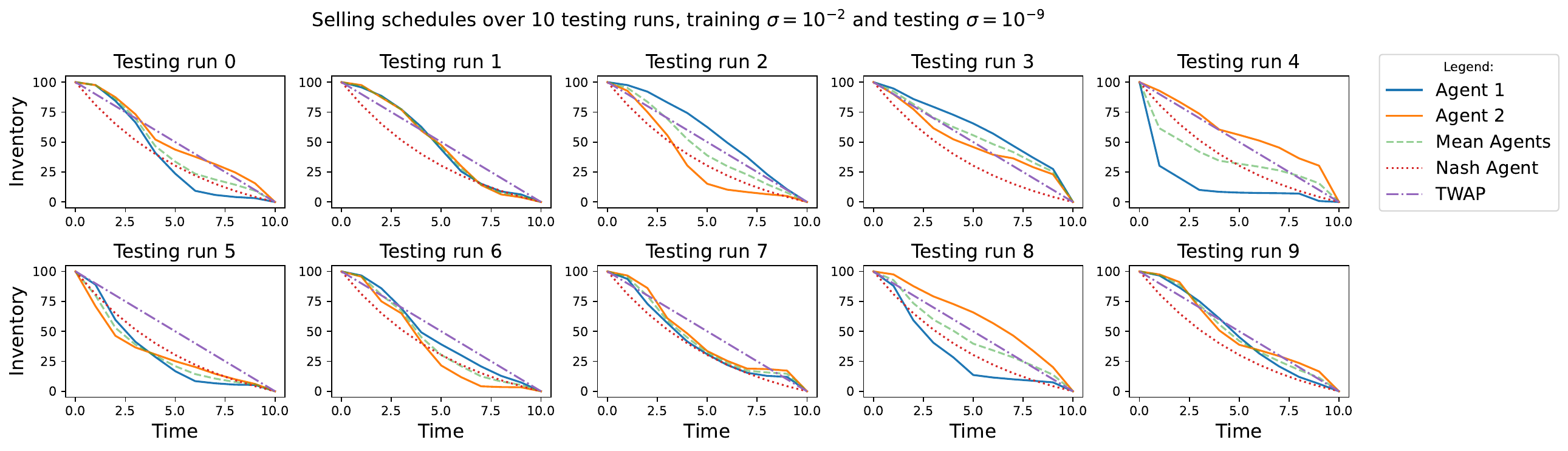}
    \caption{Optimal execution strategies for $10$ testing runs of $2,500$ iterations, training $\sigma = 10^{-9}$ and testing $\sigma = 10^{-2}$.}   \label{fig:schedule_mix_0_test_1}
\end{figure}

\subsubsection{Training with \textit{zero noise} and testing with \textit{large noise}}

When the agents are trained in an environment where $\sigma = 10^{-9}$ and the weights of this training are used in a testing environment with $\sigma = 10^{-2}$, we find that the distribution of the $IS$s per iteration is similar to the one that would be obtained in a both testing and training scenario with $\sigma=10^{-2}$ (see  Figure~\ref{fig:scatter_mix_0_test_1}). The centroids are still mostly distributed in the second, third and fourth quadrant, although some iterations and even a centroid, might end up in the first quadrant as in the $\sigma = 10^{-9}$ case. In general, it can be seen how the centroids mostly lie in the, {supra-competitive} area of the graph, i.e in the square between the Pareto-optimal $IS$ and the Nash equilibrium, pointing out at the fact that supra-competitive cost outcomes and strategies are still attainable even when the training dynamics are misspecified with respect to the testing ones.

Looking at Figure~\ref{fig:schedule_mix_0_test_1}, we notice that the selling schedules are mostly intertwined, i.e. rarely the agents trade at the same rate. Traders can still be slow or fast depending on the rate adopted by the other agent. Comparing Figure~\ref{fig:schedule_mix_0_test_1} with Figure~\ref{fig:scatter_mix_0_test_1} we see that, as in the correctly specified case, the agent that trades faster gets the lower cost, while the slow trader achieves a larger $IS$.

\subsubsection{Training with \textit{large noise} and testing with \textit{zero noise}}

\begin{figure}[ht]
    \centering
    \includegraphics[scale=0.6]{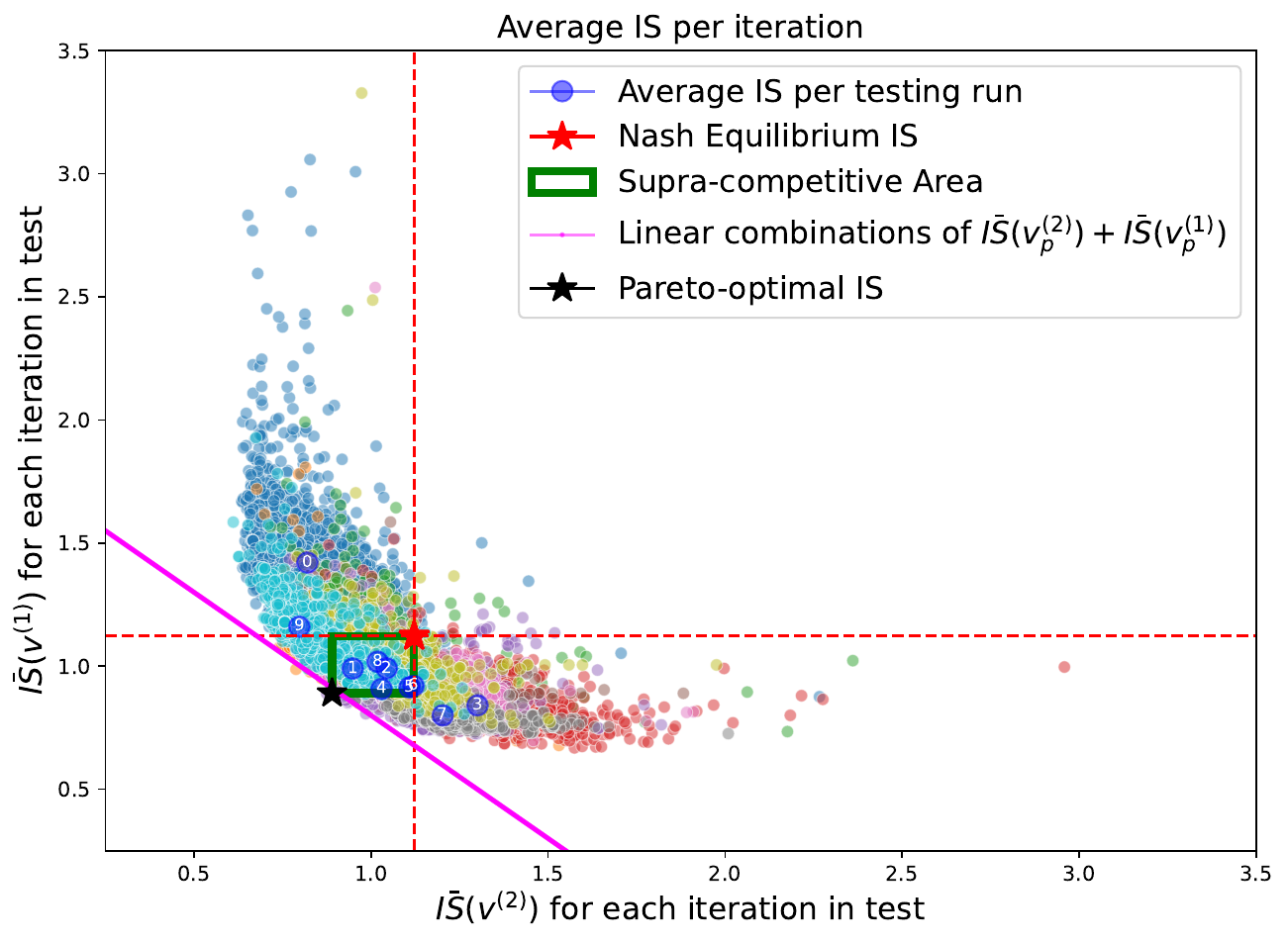}
    \caption{IS scatter for $20$ testing runs of $2,500$ iterations, training $\sigma = 10^{-2} $ and testing $\sigma = 10^{-9}$.}
    \label{fig:scatter_mix_1_test_0}
\end{figure}
\begin{figure}[ht]
    \centering
    \includegraphics[scale=0.4]{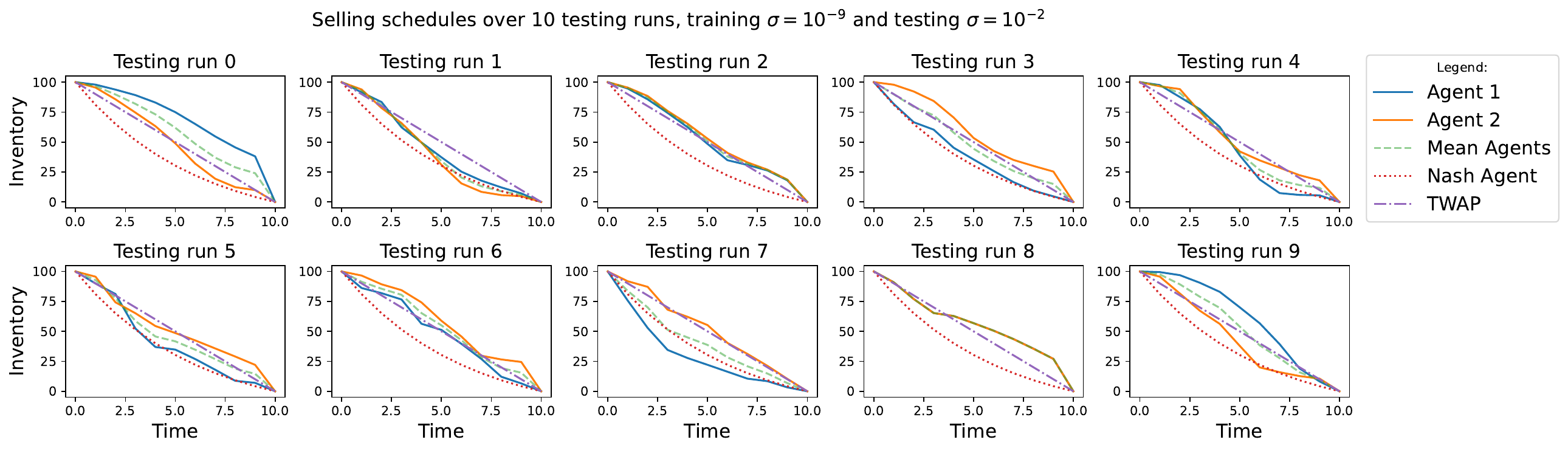}
    \caption{Optimal execution strategies for $10$ testing runs of $2,500$ iterations, training $\sigma =  10^{-2} $ and testing $\sigma = 10^{-9}$.}
    \label{fig:schedule_mix_1_test_0}
\end{figure}

Figure~\ref{fig:scatter_mix_1_test_0} shows the result of the experiment where agents are trained in an environment where $\sigma = 10^{-2}$ and tested in a market with $\sigma =  10^{-9}$. We observe that the $IS$s  are distributed in a way similar to the case where the agents were both tested and trained in an environment with $\sigma = 10^{-9}$. Still, the centroids lie in all but the first quadrant, and the majority of them lies in the supra-competitive area, showing how even with misspecified dynamics, supra-competitive strategies naturally arise in this game. The trading schedules adopted show the same kind of fast-slow trading behaviour between the agents, where still the slower trader has to pay higher costs in terms of $IS$ (see Figure~\ref{fig:schedule_mix_1_test_0}).

Finally, irrespective of the encountered levels of volatility, it seems that what is more important is the market environment experienced during the training phase, thus the selling schedule implemented in a high volatility scenario with DDQN weights coming from a low volatility scenario are remarkably similar to the one found with both training and testing with $\sigma = 10^{-9}$, and vice versa when $\sigma = 10^{-2}$ is used in training and the low volatility is encountered in the test. It might be concluded that once that the agents learn how to adopt a supra-competitive behaviour in one volatility regime, they still behave supra-competitively even if they are dealing with another volatility regime.

\section{{Real tacit collusion or supra-competitive solution?}}
{
According to \cite{harrington2018developing}, <<Collusion is when firms use strategies that embody a reward-punishment scheme that rewards a firm for abiding by the supra-competitive outcome and punishes it for departing from it>>.
Are we looking at real collusive behaviour then? Judging by the average trading strategies in Figure~\ref{fig:trading_tot} one can infer that the agents have learnt to adopt a strategy compatible with a supra-competitive equilibrium, very close to the Pareto optimal strategy. Hence, from a `global' point of view, the strategies suggest a cooperative behaviour by the agents that is directed at obtaining lower costs if compared to the Nash equilibrium strategy, as this strategy jointly minimises the costs for \textit{both} the agents collectively.
But this alone is not sufficient to talk about collusion, although tacit. 
\\
If we consider a more `local' point of view, run-wise we see how
the average of the testing iterations results in centroids that do distribute along the Pareto efficient set of solutions as in Figure~\ref{fig:scatt_tot}, and this depends on the strategies that, as we saw, show different speeds of trading per agent. In fact, comparing the centroids in Figure \ref{fig:scatter_001} with the average trading patterns in Figure~\ref{fig:trading_sigma001}, it appears that the points in upper (lower) branch of the Pareto front correspond to cases where agent 1 trades slowly (fast) achieving a higher (lower) cost, while the opposite is true for agent 2.
Thus, one can conjecture that the different speeds at which the agents trade in each iteration are a response to that deviation from the Pareto-optimal strategy, as they lie over the efficient frontier that is the set of solutions to the multi-objective optimisation problem.}

{Is it possible to interpret these deviations from the collusive strategy as a form of punishment? In our view, there is no simple answer to this question. On the one hand, considering the peculiar nature of the game itself, which is repeated and iterated because of the learning process, it is not possible to look at short deviations from the equilibrium and an eventual reversion to supra-competitive strategies, in the long run, because of a punishment mechanism. On the other hand, one can interpret the long run as a repetition of what was learned and observed in terms of the average strategy, which is in fact supra-competitive, as seen in Figure~\ref{fig:trading_tot}.
In summary, we used the definition of collusion framed in \cite{cont2022dynamics}. However, in order to be conservative we cannot conclude that the observed behaviour is a tacit collusion in the sense of \cite{harrington2018developing}. Clearly more analyses are required to answer this question, possibly considering longer iterations allowing to identify evidences of punishment by shocking the action of one of the agent as in \cite{calvano2020artificial}. We leave this interesting investigation for future research.
\\
As a final note, we highlight that such behaviour could cause welfare reduction for market participants, in terms of a distorted price formation process.}

{In fact, it is interesting to notice that the observed trading behaviour might have adverse effects on other market participants' welfare. In the Nash equilibrium the two agents trade more at the beginning of the time interval, the way information is incorporated  into the price is relatively fast. On the contrary, when the two agents are colluding they spread more evenly the execution over the time window cooperatively. In this way, private information is incorporated more slowly. Ceteris paribus, this has a negative impact on other market participants by distorting the price formation process, since the mid-price would not accurately describe the market demand and offer mechanism for the stock under execution. This would negatively affect the price formation process for other traders and eventually reduce the welfare of traders who trade based on information available on the market.}

{In conclusion, even if there no direct evidence of a reward-punishment scheme adopted by the agents, there are possibly negative consequences in terms of welfare at the expense of other market participants, while the agents that are doing optimal execution would benefit by cost reductions.
}

\section{Conclusions}

In this paper we have studied how {supra-competitive} strategies arise in an open loop two players' optimal execution game. We first have introduced the concept of collusive Pareto optima, that is, a vector of selling strategies whose $IS$ is not dominated by the $IS$s achievable by other strategies for each iteration of the game. We furthermore showed how a Pareto-efficient set of solutions for this game exists and can be obtained as a solution to a multi-objective minimisation problem. Finally, we have shown how the Pareto-optimal strategy, that is indeed the collusion strategy in this setup, is the TWAP for risk-neutral agents. 

As our main contribution, we have developed a Double Deep Q-Learning algorithm where two agents are trained using RL. The agents were trained and tested over several different scenarios where they learn how to optimally liquidate a position in the presence of the other agent and then, in the testing phase, they deploy their strategy leveraging on what learnt in the previous phase. The different scenarios, large, moderate and low volatility, helped us to shed light on how the trading interactions on the same asset made by two different agents, who are not aware of the other competitor, give rise to {supra-competitive} and Pareto-efficient strategies, i.e. strategies with a cost lower than a Nash equilibrium but higher than a collusion. This, in turn, is due to agents who keep trading at a different speed, thus adjusting the speed of their trading based on what the other agent is doing.
Agents do not interact directly and are not aware of the other agent's trading activity, thus strategies are learnt from the information coming from the impact on the asset price in a model-agnostic fashion. 

Finally, we have studied how the agents interact when the volatility parameter in the training part is misspecified with respect to the one observed by the agents in the testing part. It turns out that the existence of supra-competitive strategies possibly consistent with a tacit collusive behaviour is still present, and thus robust with respect to settings when models parameters are time varying.

There are several possible extensions of our work. An obvious extension is to the setting where more than two agents are present and/or where more assets are liquidated, leading to a multi-asset and multi-agent market impact games, leveraging on the work done in \cite{cordoni2024instabilities}. Second, impact parameters are constant, while the problem becomes more interesting when liquidity is time-varying (see \cite{macrì2024reinforcement} for RL optimal execution with one agent). Third, we have considered a linear impact model, whereas it is known that impact coefficients are usually non-linear and might follow non-trivial intraday dynamics.  Finally, the Almgren-Chriss model postulates a permanent and fixed impact, while many empirical results point toward its transient nature (see \cite{bouchaud2009markets}). Interestingly, in this setting the Nash equilibrium of the market impact game shows price instabilities in some regime of parameters (see \cite{schied2019market}), which are similar to market manipulations. It could be interesting to study if different market manipulation practices might arise when agents are trained, as in this paper, with RL techniques. The answer would certainly also be of great interest to regulators and supervising authorities.

\section*{Acknowledgements}
The authors thank Sebastian Jaimungal for the useful discussions and insights.
AM thanks Felipe Antunes for the useful discussions and insights. FL acknowledges support from the grant PRIN2022 DD N. 104 of February 2, 2022 ``Liquidity and systemic risks in centralized and decentralized markets”, codice proposta 20227TCX5W - CUP J53D23004130006 funded by the European Union NextGenerationEU through the Piano Nazionale di Ripresa e Resilienza (PNRR).

\begin{appendices}
    \section{Further learning experiments}
    \label{app:other_experiments}
    In this appendix we briefly analyse the behaviour of the algorithm outlined in Section~\ref{sec:metodi} in two different scenarios. In the first case we analyse the behaviour of two independent agents, we consider the same scenario where both the agents have to solve the optimal execution problem in a market impact game but we let them explore following a different exploration rule. In the second case we change the scenario and train just one agent to solve the optimal execution problem while the other agent consistently plays the Nash equilibrium strategy outlined in Eq.~\eqref{eq:opt_inv_two_players}. Clearly, in this last case we expect the agent to learn a strategy similar to the Nash strategy. As clearly in this case the learning agent has no incentive to deviate given the strategy of the other player, who is -by design- following the allocation suggested in Eq.~\eqref{eq:opt_inv_two_players}.
    \paragraph{Simultaneous learning around Nash.}
    
\begin{figure}[H]
        \centering
        \includegraphics[width=\linewidth]{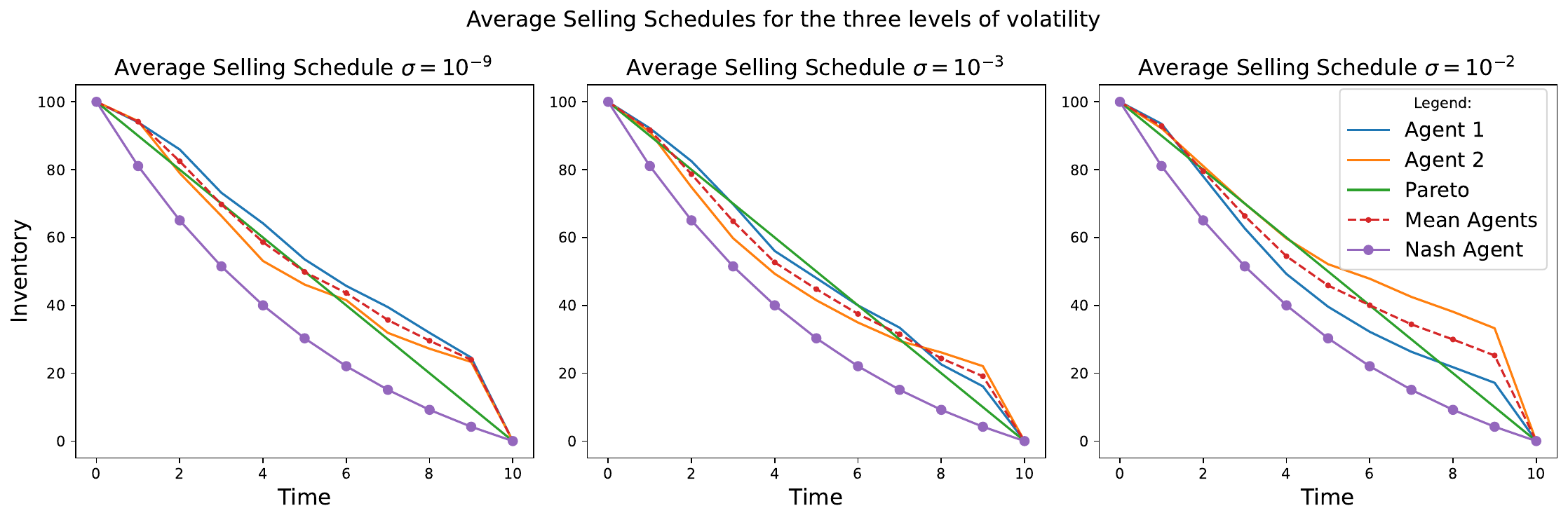}
        \caption{Average optimal execution strategy over the 10 testing runs of 2500 iterations, for all the values of $\sigma$ considered using exploration policy centred around Nash actions.}
        \label{fig:expl_nash}
    \end{figure}

    For the first set of complementary experiments, we consider agents that start exploring using the Nash equilibrium strategy, but leaving everything unchanged as in Section~\ref{sec:metodi}. To this end, we have modified the exploration policy in Eq.~\eqref{eq:action}, such that for agent $i = 1,2$ the exploration/exploitation rule is:
    \begin{align}
    \begin{split}
        &{\epsilon\in(0,1)\,\,,\,\, \zeta_i \sim\mathcal{U}(0,1)}\\
        &{v^{(i)}_{t} } = 
        \begin{cases}
             {\sim \mathcal{N}(\mu = v_t^N, \delta = |v_t^*|)\,\,\,\,\,\,\,\,\,\,\,\,\,\,\,\,, \text{if}\,\,\zeta_i \leq \epsilon}\\
            {\argmax_{v'\in[-q_t,q_{t}]}Q_{M}(g^{i}_{t},v'|\theta_{\text{main}}) \,\,\,\,\,\,\,\,, \text{else}}
        \end{cases}
    \end{split}
\end{align}
   Where \(v_t^N\) is the action for the Nash equilibrium strategy at time step \(t\). 
   We experiment over 10 testing runs of 2500 iterations, for all the values of $\sigma$. 
   We report the results in Fig.~\ref{fig:expl_nash}. It can be noticed how the agents recover strategies similar to those observed in Fig.~\ref{fig:trading_tot} in Sec.~\ref{sec:results}, hence strategies that -on average- fluctuate over the TWAP strategy and away from the Nash equilibrium strategy that was the initial exploration strategy.

    \paragraph{Single learning in a multi-player setup.}
    
\begin{figure}[H]
    \centering
    \includegraphics[width=\linewidth]{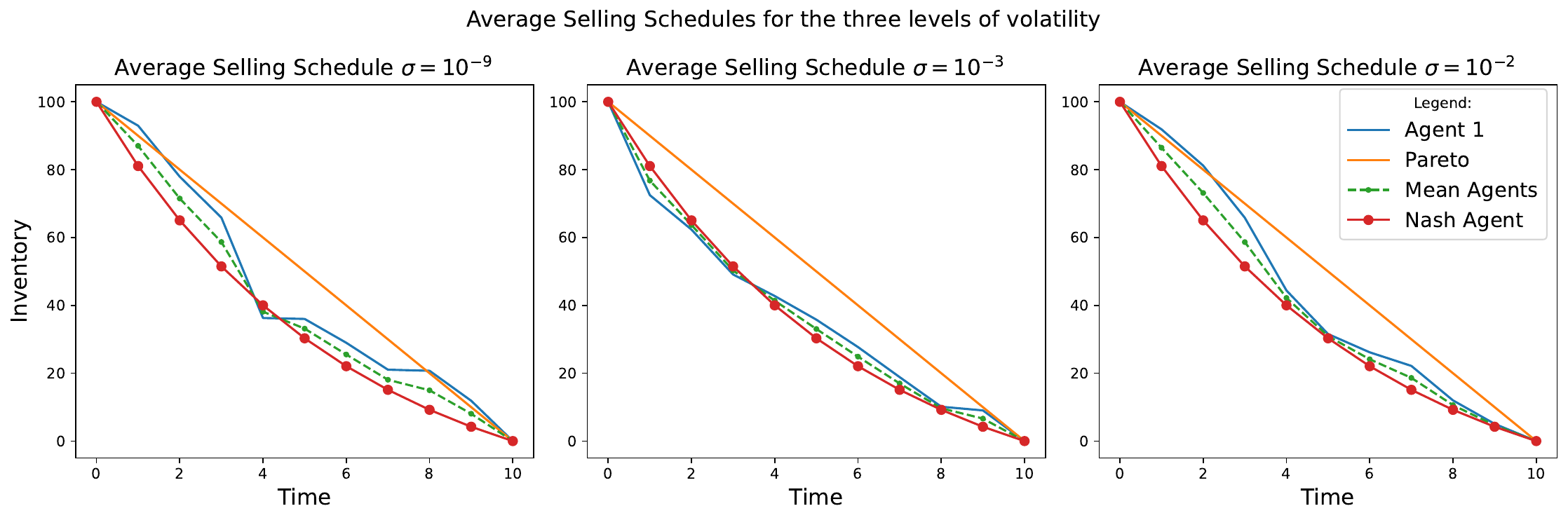}
    \caption{Average optimal execution strategy over a run of 2500 iterations, for all the values of $\sigma$ considered. Agent 2 consistently employs the Nash strategy while Agent 1 is being trained.}
    \label{fig:agent_nash_ref_2}
\end{figure}

    We consider the case of an agent who is playing the Nash equilibrium solution, while the other is learning according to the rule reported in Eq.~\eqref{eq:action}. We find that the latter agent is able to learn a trading strategy that overlaps with the true Nash equilibrium strategy, as reported in Fig.~\ref{fig:agent_nash_ref_2}. In this case we expect the agent to find this strategy as the Nash equilibrium strategy is the optimal -best response- strategy in this case.

\end{appendices}

\printbibliography

\end{document}